\documentclass{article}
\usepackage{amsmath,amssymb,latexsym}
\usepackage{stmaryrd}
\usepackage{amsfonts}
\usepackage{bm,xspace}
\usepackage{bibunits}
\usepackage{bbm}
\usepackage{url}
\usepackage{enumerate}
\usepackage[pdftex]{graphicx}
\usepackage{multirow}
\usepackage{subfigure}
\usepackage{array}
\usepackage[usenames]{color}

\definecolor{lgreen} {RGB}{180,210,100}
\definecolor{dblue}  {RGB}{20,66,129}
\definecolor{ddblue} {RGB}{11,36,69}
\definecolor{lred}   {RGB}{220,0,0}
\definecolor{nred}   {RGB}{224,0,0}
\definecolor{norange}{RGB}{230,120,20}
\definecolor{nyellow}{RGB}{255,221,0}
\definecolor{ngreen} {RGB}{98,158,31}
\definecolor{dgreen} {RGB}{78,138,21}
\definecolor{nblue}  {RGB}{28,130,185}
\definecolor{jblue}  {RGB}{20,50,100}

\newcommand{\anglebr}[1]{{\left\langle{#1}\right\rangle}}
\newcommand{\setbr}[1]{{\left\{{#1}\right\}}}
\newcommand{\paranbr}[1]{{\left({#1}\right)}}
\newcommand{\squarebr}[1]{{\left[{#1}\right]}}
\newcommand{\Gaussian}[1]{{\sN\left({#1}\right)}}
\newcommand{\abs}[1]{{\left|{#1}\right|}}
\newcommand{\norm}[1]{{\left\|{#1}\right\|}}
\newcommand{\bits}[1]{\ensuremath{\llbracket#1\rrbracket}}
\newcommand{\f}{\bmath{f}}

\def \err{{\varepsilon}}
 \newcommand{\xmath}[1]{{\ensuremath{#1}\xspace}}
 \newcommand{\bmath}[1]{\xmath{\bm{#1}}}

 \def \S{{S}}

 \def \Dbox{{\mathcal D}_{\xmath{{\rm BOX}}}}

  \def \l{\ell}

  \def \bA{\bmath{A}}
 \def \bx{\bmath{x}}

 \def \bhf{\bmath{\widehat{f}}}
 \def \btf{\bmath{\widetilde{f}}}
 
 \def \by{\bmath{y}}
 \def \bn{\bmath{n}}
 \def \bz{\bmath{z}}
 \def \bp{\bmath{p}}
 
 \def \bM{\bmath{M}}

 \newcommand{\tf}{{\widetilde{f}}}

\def \hR{\widehat{R}}

\def \hf{\widehat{f}}

\def \hS{\widehat{S}}

\def \hp{\widehat{p}}
\def \tS{\widetilde{S}}

\def \pen{\xmath{{{\rm pen}}}}
\def \penN{\pen_N}

\def \sN{{\mathcal N}}
\def \sG{{\mathcal G}}

\def \sS{\mathcal{S}}

\def\ith{{\xmath{i^{{\rm th}}}}\xspace}

\newcommand{\Smstar}{S_m^*}

\newcommand{\hlambda}{\widehat{\lambda}}

 \def\argmin{\mathop{\rm arg\,min}}
 
 \newcommand{\reals}{{\xmath{\mathbb{R}}}}
 
 \newcommand{\expect}{{\xmath{\mathbb{E}}}}
 \newcommand{\expectbr}[1]{\expect\left[{#1}\right]}
\newcommand{\prob}{{\xmath{\mathbb{P}}}}
 \newcommand{\ind}[1]{{\xmath{\mathbb{I}_{\left\{#1\right\}}}}}
 \newcommand{\eref}[1]{(\ref{#1})}

\newcommand{\ty}{\widetilde{y}}
\newcommand{\bty}{\bmath{\ty}}

\newcommand{\bI}{\bmath{I}}

\newcommand{\dou}{\partial}

\newcommand{\tA}{\widetilde{A}}
\newcommand{\btA}{\bmath{\tA}}
\newcommand{\ones}{\mathbbm{1}}
\newcommand{\zeros}{\bmath{0}}

\newcommand{\Sstar}{S^{*}}

\newcommand{\sSM}{\sS_M}
\newcommand{\tSN}{\tS_N}
\newcommand{\SstarN}{S^{*}_N}
\newcommand{\SN}{S_N}
\newcommand{\hSN}{\hS_N}
\newcommand{\RiskN}{R_N}
\newcommand{\Riskc}{R}
\newcommand{\hRiskN}{\hR_N}
\newcommand{\errN}{\err_N}

\newcommand{\tRiskN}{\widetilde{R}_N}

\newcommand{\tgamma}{\widetilde{\gamma}}
\newcommand{\hgamma}{\widehat{\gamma}}

\newcommand{\btz}{\widetilde{\bz}}

\newcommand{\ta}{\widetilde{a}}
 \newcommand{\vol}[1]{\textrm{vol}\left(#1\right)}
\newcommand{\widesim}[2][1.5]{
  \mathrel{\overset{#2}{\scalebox{#1}[1]{$\sim$}}}
}
\def\deq{\stackrel{\scriptscriptstyle\triangle}{=}}

\newcommand{\wub}[1]{{#1}}
\newcommand{\kk}[1]{{#1}}
\newtheorem{theorem}{Theorem}

\newtheorem{corollary}{Corollary}

\newcounter{subeq}

\newcommand{\done}[1]{{\textcolor{black}{#1}}}

\title{Level set estimation from projection measurements: Performance guarantees and fast computation \thanks{Kalyani Krishnamurthy and Rebecca Willett were supported by NGA Award No. HM1582-10-1-0002 SUB \#1-P3130108 and AFRL Grant No. FA8650-07-D-1221. Waheed U. Bajwa was supported in part by the NSF under grant CCF-1218942.}}
\author{Kalyani Krishnamurthy\footnotemark[2]\thanks{Department of Electrical and Computer Engineering, Duke University, Durham, NC 27708 (kk63@duke.edu, willett@duke.edu)} \and Waheed U. Bajwa\footnotemark[3]\thanks{Department of Electrical and Computer Engineering, Rutgers University, Piscataway, NJ 08854 (waheed.bajwa@rutgers.edu)} \and Rebecca Willett\footnotemark[2]
}

\begin{document}
\maketitle
\sloppypar
\begin{abstract}
Estimation of the level set of a function (i.e., regions where the function
exceeds some value) is an important problem with applications in digital
elevation mapping, medical imaging, \wub{astronomy, etc.} In many
applications, \wub{the function of interest is not observed directly. Rather,
it is acquired through (linear) projection measurements,} such as tomographic
projections, interferometric measurements, coded-aperture measurements,
\wub{and random} projections associated with compressed sensing. This paper
describes a \wub{new methodology for rapid} and accurate estimation of the
level set from such projection measurements. \wub{The key defining
characteristic of the proposed method, \kk{called the projective level set estimator,} is its ability to estimate the level
set from projection measurements without an intermediate reconstruction step.
This leads} to significantly faster computation relative to \wub{heuristic}
``plug-in" methods \wub{that} first estimate the function, typically with an
iterative algorithm, and then threshold the result. \wub{The paper also
includes a rigorous theoretical analysis of the proposed method, which
utilizes 
\done{results from the literature on concentration of measure} and
characterizes the estimator's performance in terms of geometry of the
measurement operator and \kk{$\ell_1$}-norm of the discretized function.}
%
\end{abstract}

\section{\wub{Introduction}}
\kk{Level set estimation is the process of using indirect
observations of a function $f$ defined on the unit hypercube $[0,1]^d$ to
estimate the region(s) where $f$ exceeds some critical value $\gamma$; i.e.,
$\Sstar \deq \left\{ x \in [0,1]^d : f(x) > \gamma \right\}$.}
{Accurate and efficient level set estimation plays a crucial role in a
variety of scientific and engineering tasks, including the localization of
``hot spots'' signifying tumors in medical imaging
\cite{li2006fast,harmany2008controlling}, significant photon sources in
astronomy \cite{clusterAnalysisAstronomy}, \wub{and} strong reflectors in
remote sensing \cite{ayed2005multiregion,marques2009target}.


\kk{In this paper, we consider making observations of the form $\by = \bA \f +\bn,$ where \kk{$\f$ is a discretized version of $f$}, $\bA$ is a
\wub{(discrete)} linear operator that may not be invertible, and $\bn$ is additive noise that corrupts our observations.}
\done{For instance,
$\by$ might correspond to tomographic projections in tomography
\cite{tomographyBook,lewitt1979processing,huang1991optical}, interferometric measurements in radar interferometry
\cite{rosen2000synthetic}, multiple blurred, low-resolution, dithered
snapshots in astronomy \cite{puetter2005digital}, or \wub{random} projections
in compressed sensing systems
\cite{spm:cs08,CS:candes1,CS:candes2,CS:donoho,riceCamera}.}
Our goal in this \wub{$\by = \bA \f +\bn$} setting is to perform level set
estimation \done{of the continuous-domain function $f$} {\em without} an
intermediate step involving time-consuming reconstruction of $\f$. There are
two reasons for this\wub{.} First, level set estimation without
reconstruction of $\f$ would allow \wub{design of sequential measurement
schemes optimally adapted to the function of
interest.} 
For instance, in tomography we would like to estimate \wub{the level set,
$\Sstar$,} quickly from \wub{an initial set of} observations so that
\wub{additional observations focused on $\Sstar$} can be collected
immediately, resulting in an overall low radiation dose
\cite{kyrieleisregion,maa§2011new,JunMa2010}. Some recent \wub{works}
\cite{haupt2009distilled, haupt2009compressive} \wub{have provided}
theoretical \wub{characterizations} of the significant benefits associated
with certain sequential measurement schemes; the \wub{method} proposed in
this paper may facilitate the use of such schemes in time-sensitive or
computational-resource limited applications. Second, ``plug-in'' approaches
that estimate $\f$ and threshold the estimate $\bhf$ to extract $\Sstar$ are
notoriously difficult to characterize; the performance hinges upon the
statistics of the \wub{estimation error} $\bhf-\f$, which for most
reconstruction methods are unknown (with the possible exception of the first
moment). More generally, reconstruction methods aim to minimize the total
error, integrated or averaged spatially over the entire function. This does
little to control the error at specific locations of interest, such as in the
vicinity of the level set boundary. {Finally, the Vapnik Principle
\cite{vapnik2000nature} states that one should never solve a complex problem
as an intermediate step towards solving a simple problem.}

\subsection{Problem formulation}
\label{Sec:ProbForm}
In
this work, we observe samples of a function $f$ supported on $[0,1]^d$ of the form
\begin{align}
\by = \bA \f + \bn \in \reals^K
\label{eqn:probDesc}
\end{align}
where
\begin{itemize}
\item $\bA \in \reals^{K \times N}$ is a \wub{linear}
operator that is assumed to be known with $K$ often less than $N$,
\item $\f \in \reals^N$ corresponds to integration samples of $f$; i.e.,
\begin{align}
f_{i} = \frac{1}{\vol{C_i}}\int_{C_i}f(x) dx \label{eqn:discretization}
\end{align}
for $i = 0,1,\ldots,N-1$\wub{,} where the cells $C_i$'s are obtained by
partitioning $[0,1]^d$ \wub{into} nonoverlapping hypercubes such that
each $C_i$ has sidelength $N^{-1/d}$ and volume $1/N$, and
\item \done{$\bn \in  \reals^K$ denotes the additive measurement noise,
    which is assumed to be zero-mean, subGaussian white noise in our
    case; i.e., $n_i \widesim{\mathrm{i.i.d.}} \text{Sub}(c_s)$ is a
    zero-mean, subGaussian random variable, defined by the condition
    $\paranbr{\expectbr{\abs{n_i}}^p}^{1/p} \leq c_s \sqrt{p}$ for $p
    \geq 1$}.\footnote{\done{Note that the subGaussian noise assumption
    subsumes the usual assumption of Gaussian noise; in particular,
    Gaussian random variables and bounded random variables fall under the
    category of subGaussian random variables
    \cite{rudelson2013lecture}.}}
\end{itemize}
\done{We assume without loss of generality that the columns of $\bA$ have
unit $\ell_2$ norms and consider $N$ to be dyadic (power of two).} A
$\gamma$-level set in this discrete setting can be written as \kk{$\SstarN =
\{i : f_i > \gamma\}$ where the subscript $N$ signifies that the
\done{discrete-domain} level set is a function of the $N$-dimensional
discrete signal $\f$.\kk{\footnote{\kk{In this work, we adopt the terminology
of ``function" for the continuous-domain $f$ and ``signal" for its discrete
counterpart $\f$.}}}
Throughout this paper the dependencies
of the continuous-domain level set $\Sstar$ and the discrete-domain level set $\SstarN$ on $\gamma$ are implicit.}

\done{Our main goal is to estimate the continuous-domain level set $\Sstar$
from discrete measurements $\by$ \emph{without reconstructing} the underlying
\kk{signal} $\f$. In the discussion that follows, we propose a level set
estimation method to estimate the discrete-domain level set $\SstarN$
directly from $\by$ and show that $\SstarN \longrightarrow \Sstar$ as $N
\longrightarrow \infty$ in Sec.~\ref{Sec:CSLSEst}.} {Similar to
\cite{willett:levelset}, \wub{the error metric used to measure the closeness}
between $\SstarN$ and a candidate estimate $\S$ \wub{is defined} as
\begin{equation}
\errN{\paranbr{\S,\SstarN}} = \frac{1}{N}\sum_{i \in \Delta(\SstarN\!,\S)} |\gamma - f_i|
\label{eqn:objective}
\end{equation}
where \wub{$\Delta(\SstarN\!,\S) \deq \{i \in (\SstarN \setminus \S) \cup (\S
\setminus \SstarN)\}$ denotes the symmetric set difference between $\S$ and
$\SstarN$.}
Note that \eref{eqn:objective} \kk{can be interpreted as an empirical, weighted probability of
error under the counting measure} where the weights depend on the amplitude of the \kk{signal} relative to
the level set threshold $\gamma$. \done{Our error metric penalizes (a) the symmetric difference between a level set estimate $S$ and the true level set $\SstarN$, and (b) the errors along regions of the level set boundary corresponding to abrupt intensity variations more than the regions where the intensity varies smoothly. This performance measure is ideally suited for the level set estimation problem since, in many applications such as localizing hot-spots signifying tumor in biomedical imaging, it is more desirable for an algorithm to accurately localize regions with sharp intensity variations. }

\wub{Instead of working directly with the error metric, we make use of the
{\em risk} of a candidate set $\S$, defined as}
\begin{equation}
\RiskN(\S) \deq \frac{1}{N}\sum_{i} {\ell_i(\S)}
\label{eq:risk}
\end{equation}
where
\begin{align}
\ell_i(\S) \deq \left(\gamma - f_i\right) \left[\ind{i\in \S}-\ind{i
    \notin \S} \right]
    \label{eqn:lossFn}
\end{align}
is the {\em loss} function and $\ind{E} = 1$ if event $E$ is true and $0$
otherwise. The loss function in \eref{eqn:lossFn} measures the distance
between the \kk{signal value} at location $i$, $f_i$, and the threshold, $\gamma$, and
weights this distance by $-1$ or $1$ according to whether $i\in \S$ or not.
The loss function $\ell_i(\SN)$ is positive if $i \in \Delta(\SstarN\!,\S)$ and
is negative otherwise. To see this, observe that for all \wub{$i \in \SstarN
\setminus \S$}, $\paranbr{\gamma-f_i} \leq 0$ and $\left[\ind{i\in \S}-\ind{i
    \notin \S} \right] = -1$. A similar explanation holds for all \wub{$i \in \S \setminus \SstarN$} as well. Note that \wub{the risk is
    related to the error metric defined in \eqref{eqn:objective} by virtue of
    the fact that}
\begin{align}
\RiskN(\S) - \RiskN(\SstarN) &= \frac{1}{N} \sum_{i} \left(\gamma - f_i\right) \paranbr{\left[\ind{i\in \S}-\ind{i
    \notin \S} \right] -  \left[\ind{i\in \SstarN}-\ind{i
    \notin \SstarN} \right]} \nonumber \\
&= \frac{2}{N} \sum_{i \in \Delta(\SstarN,\S)} \abs{\gamma-f_i} = 2\errN{\paranbr{\S,\SstarN}}. \label{eqn:excessRiskdiscrete}
\end{align}
Finding an estimator that minimizes the \wub{\emph{excess risk error}}
$\errN{\paranbr{\S,\SstarN}}$ is thus equivalent to finding an estimator that
minimizes $\RiskN(\S)$ since $\RiskN(\SstarN)$ is simply a constant with respect to $\S$.}

{This paper presents an optimization problem for
choosing an estimate of $\SstarN$ from the data $\by$
and theoretical characterization of $\errN{\paranbr{\SN,\SstarN}}$ when $\f$
consists of samples of a piecewise smooth function.}

 \section{Our contribution and relation with previous work}
 In this work, we demonstrate that, subject to certain conditions on $\bA$ and the $\ell_1$ norm of $\f$, the level set $\Sstar$ can be
estimated quickly and accurately via $\SstarN$ without first reconstructing $\f$.
 For $\bA = \bI$, \cite{willett:levelset} provides minimax optimal, tree-based level set
estimation techniques to extract $\Sstar$ from noisy observations $\by =
\f+\bn \in \reals^N$ without estimating $\f$. We cannot directly apply those results to our problem since $\bA \neq \bI$.
\wub{Instead}, we draw on the key idea of constructing \wub{\emph{proxy
observations}}
\begin{align}
\bz = \bA^T\by = \f + \underbrace{\left(\bA^T\bA-\bI \right)\f + \bA^T\bn}_{\bn'} \label{eqn:proxy}
\end{align}
from the \wub{literature on support detection of sparse signals (see, e.g.,
\cite{bajwa:jc10, fletcher:tit09, wasserman})} and then exploit some of the
important insights from \cite{willett:levelset} to address our problem. A part of this work was previously published in \cite{krishnamurthy2011fast}. This work, however, significantly expands on the previous work and presents new and tighter theoretical bounds and extensive simulation experiments.

Before we present our estimation method, we discuss prior work on level set estimation and sparse support detection.

\subsection{Previous work on level set estimation}\label{ssec:previous}
{Large volumes of research have been dedicated to the problem of estimating
level sets of an unknown density or a regression function $f$ from its noisy
measurements by either using plug-in estimators that find level sets of
estimates of $f$
\cite{moreno2003total,cuevas2006plug,rigollet2006fast,singh2009adaptive,mason2009asymptotic}
or direct methods that do not involve an intermediate reconstruction step
\cite{tsybakovDensity,scott2006minimax,willett:levelset,scott2007regression,clayAnomaly}.
Plug-in methods are easy to implement and in some cases lead to theoretical
results on consistency and convergence based on some smoothness assumptions
on the function of interest. For instance,
\cite{moreno2003total,cuevas2006plug,rigollet2006fast} propose plug-in
methods based on kernel estimators and show that they exhibit fast rates of
convergence. Mason and Polonik \cite{mason2009asymptotic} derive the
asymptotic normality of the symmetric difference between a true level set and
an estimate derived using a kernel density based plug-in estimator. Singh,
Scott and Nowak \cite{singh2009adaptive} propose a plug-in method based on a
regular histogram partition that minimizes the Hausdorff distance between the
true and the estimated level sets. They also demonstrate that the proposed
method adapts to unknown regularity parameters and achieves near minimax
optimality on a wide variety of density function classes.

\done{In the specific $\by = \bA \f + \bn$ case studied in this paper, a
number of plug-in methods can be proposed by exploiting the vast literature
on ill-posed inverse problems \cite{Kaipio.Somersalo.Book2005}. Two popular
and computationally simple methods in this regard are the truncated singular
value decomposition (TSVD) (also known as the pseudo-inverse solution) and
Tikhonov regularization. While both these methods lead to fast plug-in
approaches to level set estimation, essentially involving first an estimation
of $\f$ from $\by$ and then thresholding of the resulting estimate, we do not
expect these approaches to perform well in practice. This is because both
TSVD and Tikhonov regularization focus on ``minimum-energy solutions,'' which
effectively involves projecting $\by$ onto the principal subspace of $\bA$.
In the case of underdetermined $\bA$, however, sparse signal processing
research in the last decade or so has established the suboptimal nature of
such ``subspace approaches'' to ill-posed inverse problems \cite{Lu.Do.ITSP2008}. Instead, the
state-of-the-art in ill-posed linear inverse problems with an underdetermined
$\bA$ involves projecting $\by$ onto a ``union of subspaces''
\cite{Duarte.Eldar.ITSP2011}, accomplished through the use of either
total-variation (TV) regularization \cite{wang2008new,rudin1992nonlinear} or
$\ell_1$ regularization \cite{twist}.
%
}

\done{While the aforementioned plug-in approaches to level set estimation
seem attractive, they solve a much harder problem as an intermediate step to
solving a set estimation problem---a problem that is simpler than function
estimation.} Vapnik's principle stated earlier, together with the minimax
convergence results shown in the context of classification problems in
\cite{yang} tell us that plug-in methods are often suboptimal to direct
estimation methods. As a result, in our work, we focus on direct set
estimation strategies.}

{Several researchers have considered direct set estimation methods for the case $\bA = \bI$.
In \cite{tsybakovDensity}, Tsybakov proposes a direct density level set estimation method that finds piecewise polynomial estimators of the true level set and achieves optimal minimax rates of convergence.
 The estimation method in \cite{tsybakovDensity} is hard to compute and cannot be directly extended to our problem where $\bA \neq \bI$.
In \cite{scott2007regression}, the authors show the theoretical and practical
advantages of reducing a regression level set estimation problem to a
cost-sensitive classification problem.
Previous work by one of the coauthors
\cite{willett:levelset} draws on the relationship between classification and level set estimation frameworks, and proposes a set estimation
method based on dyadic decision trees by exploiting some of the ideas from
\cite{scott2006minimax}.
A closely related work is the estimation of minimum volume sets such that their masses
are at least greater than some specified $\gamma$ \cite{clayAnomaly}. In that
work, the authors discuss tree-based techniques and provide universal consistency
results and rates of convergence.

{We briefly review the basic idea in \cite{willett:levelset} on which our set estimation strategy is built upon. The goal in that work was to design
  an estimator of the form
  $$\hS = \argmin_{S \in \sSM} \hRiskN(S) + \pen(S),$$
  where $\sSM$ is a class of candidate estimates,
  $\hR_N$ is an empirical measure of the estimator risk
  based on $N$ noisy observations of the \kk{signal} $\f$, and $\pen(\cdot)$ is
  a regularization term which penalizes improbable level sets. That work
  described choices for $\hRiskN$, \wub{$\pen(\cdot)$}, and $\sSM$ \wub{that}
  made $\hS$ rapidly computable and minimax optimal for a large class
  of level set problems. \wub{Specifically, it derived a regularizer
  $\pen(\cdot)$ using Hoeffding's inequality} \done{for bounded random variables \cite{hoeffding}} and }
{developed a dyadic tree-based framework to obtain $\hS$. Trees were utilized
for a couple of reasons. First, they both restricted and structured the space
of potential estimators in a way that allowed the global optimum to be both
rapidly computable and very close to the best possible (not necessarily
tree-based) estimator.  Second, they allowed the estimator selection
criterion to be spatially adaptive, which was critical for the formation of
provably optimal estimators. \wub{Note that while we intend to build upon the
insights developed in \cite{willett:levelset}, an extension of those
techniques to the case of proxy observations in \eqref{eqn:proxy} is made
nontrivial because of two reasons. First, the \emph{effective noise} $\bn'$
is nonzero mean because of the presence of $\left(\bA^T\bA-\bI \right)\f$.
Second, and most importantly, $\bn'$ is \emph{correlated} due to the
non-unitary nature of $\bA$, which prohibits the use of \done{canonical
Hoeffding's inequality \cite{hoeffding}} for characterization of the penalty
term.
%
}

\subsection{Relationship with previous work on sparse support detection}
\kk{{Sparse support detection is the problem of detecting a set of locations $\SstarN = \setbr{i: f_i \neq 0}$ corresponding to a discrete signal $\f \in \reals^N$, given observations of the form in \eref{eqn:probDesc}.
This is a special case of level set estimation and the two are equivalent if $\f$ is nonnegative and $\gamma =0$.
The idea of constructing proxy observations $\bz$ to deduce certain
  properties of the underlying $\f$ has been successfully employed in
  recent compressed sensing and statistics literature
   to solve the
  problem of support detection of a discrete $\f$ having no more than $m$ non-zero
  entries;
  see, e.g., \cite{bajwa:jc10, fletcher:tit09, wasserman,haupt2009compressive}. Specifically, it is established in \cite{bajwa:jc10} that the support of an $m$-sparse $\f$ can be reliably and quickly detected from appropriately thresholded proxy observations with overwhelming probability as long as $\bA$ satisfies a certain, easily verifiable coherence property. The success of this thresholding method stems primarily from the sparsity assumption on $\f$. However, when $\f$ is not sparse, as is the case in level set estimation, simply thresholding the proxy observations will result in numerous false positives and misses as discussed in detail in the numerical experiments in Sec.~\ref{Sec:expts}; see Figs.~\ref{fig:trueImage}, \ref{fig:trueLevelSet}, and Figs.~\ref{fig:proxynoMS} through \ref{fig:riskOptimalthrnoMS}. These results clearly suggest that we cannot simply use a support detection algorithm and an optimally chosen threshold to achieve an accurate level set estimation.} In
  contrast, our methodology relies on a novel two-step approach that
  enables us to work with proxy observations without requiring $\f$ to
  be sparse.}

\section{Fast level set estimation from projection measurements}
\label{Sec:CSLSEst}
%
In order to extract the $\gamma$-level set of $\f$ from $\by$, we propose a
novel two-step procedure. First, we construct a proxy of $\f$ according to
\eref{eqn:proxy}\wub{, which }
allows us to arrive at the canonical signal plus noise observation model.
Next, we perform level set estimation on the proxy observations $\bz$, rather
than on $\by$, using a method similar to the one derived in
\cite{willett:levelset}.\footnote{\done{There is another equivalent
understanding of our approach to level set estimation, which helps connect it
to the classical literature on inverse problems. The proxy observations $\bz$
can be thought of as setting up the \emph{normal equations} $\bA^T \by =
\bA^T \bA \widehat{\f}$. Instead of first solving the normal equations for
one of infinitely-many $\widehat{\f}$, arising due to the underdetermined
nature of $\bA$, our approach can be construed as estimating the level set
directly from the normal equations.}} \kk{We refer to the resulting estimator
as the \emph{projective level set estimator}.} Note that for any unitary
$\bA$, $\bz$ in \eref{eqn:proxy} reduces to \wub{$\by = \f + \tilde{\bn}$
with $\tilde{\bn}$ having independent, zero-mean entries}.
However, for non-unitary $\bA$, the
proxy defined in \eref{eqn:proxy} creates a signal-dependent interference
term $\left(\bA^T\bA-\bI \right)\f$ and a zero-mean \emph{correlated} noise
term $\bA^T\bn$.
}

{\wub{Intuitively, if} we try to make a decision about each $z_i$
independently, then we would be vulnerable to noise \wub{(see, e.g.,
Figs.~\ref{fig:thrnoMS} and \ref{fig:riskOptimalthrnoMS})}. \wub{On the other
hand, if} we consider \wub{\emph{patches} ${p_j}_{s}$} of $z_i$'s,
\done{defined as groups of $s$ proxy measurements ($z_i$'s) centered around $z_j$
for $s \in \{1,\ldots,N\}$}, and force each patch to be wholly inside or
outside the level set estimate, then we increase our robustness to noise but
also increase our bias. Ideally, we want spatially adaptive patches
\wub{that} allow us to balance between an accurate approximation of the true
level set boundary and estimator variance. \wub{It is in this vein that} we
theoretically analyze the impact of $\bn'$ \wub{on the level set estimation
problem} and use \wub{our analysis} to develop a spatially-adaptive, dyadic,
tree-based level set estimation approach that adapts to \wub{both} the
interference \wub{and the correlated noise} term.}

{The algorithm we propose basically works by using $\bz$ to find a partition
of $\f$ into a collection of disjoint sets of \wub{``pixels.''} For each set,
we determine whether it is inside or outside the level set with a simple
voting procedure---i.e., we \wub{determine whether} the majority of the
$z_i$'s \wub{in the set} are greater than gamma. Thus searching for the
optimal level set estimate amounts to searching for a good partition of $\f$
and then performing empirical risk minimization\wub{, defined in \eref{eqn:empiricalRiskDefn} in the sequel,}
on that partition. We restrict our attention to partitions defined using
binary trees because they yield tractable algorithms and, in the case where
$\bA=\bI$, minimax optimality \cite{willett:levelset}.}

\begin{figure}
\centering
\includegraphics[height=2.25in]{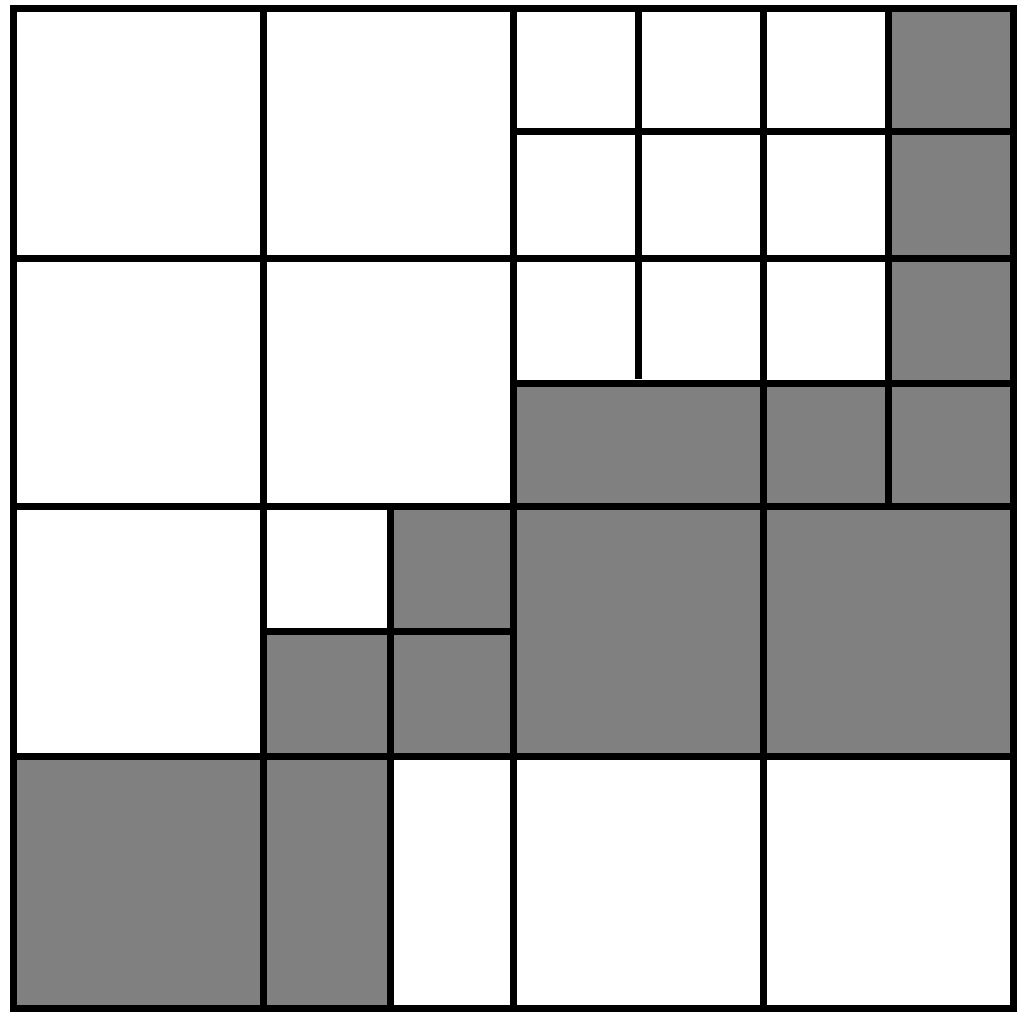}
\caption{\label{fig:levelsetPartition}An example level set estimate $\S \in \sSM$ where the domain of the underlying signal
is $[0,1]^2$. Shaded regions are estimated to be outside the level set.}
\end{figure}

{\wub{Specifically}, let $\sSM$ be a collection of candidate level set
estimates for a dyadic $M$ (i.e., $M = 2^q$ for some positive integer $q$)\wub{,} where each $\S \in \sSM$
is obtained by recursively partitioning the domain
of $\f$ in dyadic intervals.
\kk{The number of dyadic intervals along different coordinate directions is not required to be the same. In other words, each cell in the partition can potentially have different sidelengths and the sidelength of the smallest cell is $1/M$.
An estimate $\S \in \sSM$ is obtained by assigning each cell in the partition to be inside or outside of the level set. Fig.~\ref{fig:levelsetPartition} shows one such estimate in two dimensions where the shaded
regions are the partition cells that are estimated to be outside the level
set. Though we do not specify $M$ in terms of $N$ here, we derive an upper bound on $M$ as a function of $N$ that achieves a certain expected excess risk in Theorem~\ref{thm:expectedExcessRiskBounds}.
}

Given $\bz$, our goal is to find a level set estimate
\begin{align}
\tSN = \argmin_{\S \in \sSM} \RiskN(\S)-\RiskN(\SstarN) \wub{= \argmin_{\S \in \sSM} \RiskN(\S)}\label{eqn:idealEst}
\end{align}
where $\RiskN(\cdot)$ is defined in \eref{eq:risk} \wub{and the second equality
follows since $\RiskN(\SstarN)$ is a constant}. (Note that $\tSN = \SstarN$ if
$\SstarN \in \sSM$.)
Since $\f$ is unknown, \wub{$\RiskN(\S)$ cannot be computed; instead,} let us  consider
an empirical risk of the form
\begin{align}
\hRiskN(\S) = \frac{1}{N} \sum_{i=1}^N \left(\gamma-z_i\right) \left[\ind{i \in \S}-\ind{i \notin \S}\right].
\label{eqn:empiricalRiskDefn}
\end{align}
We show that finding an estimate that minimizes a penalized empirical risk results in an estimate that asymptotically approaches $\tSN$. Specifically, we find
\begin{align}
\hSN = \argmin_{\S \in \sSM} \hRiskN(\S)+\penN(\S), \label{eqn:Shat}
\end{align}
where $\penN(\S)$ is an interference-dependent penalty term that yields
$\abs{\RiskN(\hSN)-\RiskN(\tSN)} \stackrel{N \longrightarrow
\infty}{\longrightarrow} 0$ subject to certain conditions on $\bA$, which
generally require $K \longrightarrow \infty$ as $N \longrightarrow \infty$.
The penalty term plays a major role in our estimation strategy and is crucial
in finding estimates that hone in on the boundary of the level set $\SstarN$.
We thus focus on designing a spatially adaptive penalty $\penN(\S)$ that
promotes well-localized level sets with potentially non-smooth boundaries.
\kk{Let $\pi(\S)$ be the partition induced
by an estimate $\S\in \sSM$, i.e., $\pi(\S)$ is the collection of all leaves in
the estimate $\S \in \sSM$.} \kk{Fig.~\ref{fig:levelsetPartition} shows a partition induced by one of the estimates $\S \in \sSM$ where every white or gray shaded block is a leaf. \done{We assign a label $\ell(L)$ to each leaf $L$ depending on whether $L$ is in the level set ($\ell(L)=1$) or otherwise ($\ell(L)=0$).}} Then the risk of $\S$ in each of its leaf $L \in
\pi(\S)$ is given by
$$\RiskN(L) \deq \frac{1}{N} \sum_{i =1}^N \left(\gamma-f_i\right)
\left[\ind{\ell(L)=1}-\ind{\ell(L)=0}\right]\ind{i \in L}.$$
Note that $\RiskN(\S) =
\sum_{L \in \pi(\S)} \RiskN(L)$. We design a spatially adaptive penalty term by
analyzing $\RiskN(L)-\hRiskN(L)$ \wub{within} each leaf separately. To facilitate our
analysis, let us \kk{define}
$$\tRiskN(L) \deq \frac{1}{N} \sum_{i =1}^N \left(\gamma-\expectbr{z_i}\right) \left[\ind{\ell(L)=1}-\ind{\ell(L)=0}\right]\ind{i \in L}.$$ Then
\begin{align}
\abs{\RiskN(L)-\hRiskN(L)} &= \abs{\RiskN(L)-\tRiskN(L)+\tRiskN(L)-\hRiskN(L)}\nonumber \\
=&\abs{\frac{1}{N} \sum_{i=1}^N \squarebr{\left(\expectbr{z_i}-f_i\right)+\left(z_i-\expectbr{z_i}\right)}\left[\ind{\ell(L)=1}-\ind{\ell(L)=0}\right]\ind{i \in L}}\nonumber\\
\leq& \underbrace{\abs{\frac{1}{N} \sum_{i=1}^N \left(\expectbr{z_i}-f_i\right)\left[\ind{\ell(L)=1}-\ind{\ell(L)=0}\right]\ind{i \in L}}}_{T_1} \label{eqn:T1T2} \\
&+ \underbrace{\abs{\frac{1}{N} \sum_{i=1}^N\left(z_i-\expectbr{z_i}\right)\left[\ind{\ell(L)=1}-\ind{\ell(L)=0}\right]\ind{i \in L}}}_{T_2}.\nonumber
\end{align}
Note that while $T_1$ is a measure of the bias in $\bz$, $T_2$ is a measure of the concentration of $\bz$ about its mean.
Since the columns of $\bA$ are assumed to have unit $\ell_2$ norms, one can easily see from \eref{eqn:proxy} that
\begin{align}
z_i = f_{i} + \sum_{j=1, j \neq i}^{N} f_j \anglebr{\bA^{(i)},\bA^{(j)}}+ \anglebr{\bA^{(i)},\bn} \label{eqn:z_i}
\end{align}
where $\bA^{(i)}$ denotes the $\ith$ column of $\bA$ \wub{and
$\anglebr{\cdot,\cdot}$ denotes the usual innerproduct}. Since $\bA$ is
given, and $\bn$ is zero mean, the term
\begin{align}
\expectbr{z_i} - f_i = \sum_{j=1, j \neq i}^{N} f_j \anglebr{\bA^{(i)},\bA^{(j)}} \label{eqn:expect_z_i}
\end{align}
in $T_1$ is the signal-dependent interference term at the $\ith$ location due
to the signal energies at other locations. We upper bound $T_1$ by the
$\ell_1$ norm of $\f$ and the worst-case coherence of $\bA$ \kk{(defined in
the statement of Theorem \ref{thm:CSLevelSets})}, bound $T_2$
using 
\done{a Hoeffding-like inequality for a weighted sum of independent
subGaussian random variables \cite{rudelson2013lecture}}, and sum the risk in
each leaf of the estimate $\S$ to arrive at \kk{the following} result.
\begin{theorem}[Concentration of risk \wub{around the empirical risk}]
\label{thm:CSLevelSets}
\done{Suppose that the entries of noise $\bn$ are subGaussian distributed
with parameter $c_s$.} Then, for $\delta \in [0,1/2]$ and $c>0$, with probability at
least $1-2\delta$, the following holds for all $\S \in \sSM$:
\begin{align}
\abs{\RiskN(\S)-\hRiskN(\S)} \leq &\left(\frac{N-1}{N}\right)\mu(\bA) \|\f\|_{1} + \penN(\S)\label{eqn:finalRiskBound}
\end{align}
where \kk{$\norm{\f}_1 = \sum_i \abs{f_i}$ is the $\ell_1$ norm of $\f$,}
\begin{align}
\penN(\S) \deq \sum_{L \in \pi(\S)}\frac{1}{N}\sqrt{\frac{\left[\log(2/\delta)+\bits{L}\log 2\right] \done{c_s^2}\sum_{i,j \in L}\anglebr{\bA^{(i)},\bA^{(j)}}}{2c}} \label{eqn:Phi}
\end{align}
is the penalty term, $$\mu(\bA) \deq \max_{i,j \in \{1,\ldots,N\}, i \neq
j}\abs{\anglebr{\bA^{(i)},\bA^{(j)}}}$$ is the worst-case coherence of $\bA$
and \kk{$\bits{L}$ is the number of bits in a prefix code used to uniquely encode the position of a leaf $L$ in the tree. }
\end{theorem}

The proof of this theorem is provided in Section~\ref{sec:CSLevelSets}.
\kk{The above bound holds for \emph{any} prefix code $\bits{L}$. In order to
achieve the error rates in Theorem~\ref{thm:expectedExcessRiskBounds}, we use
a certain prefix code, which is discussed before the statement of
Theorem~\ref{thm:expectedExcessRiskBounds}.} Note that the bounds in
\eref{eqn:finalRiskBound} and \eref{eqn:Phi} depend on (a) the
signal-dependent interference term in \eref{eqn:proxy} through $\norm{\f}_1$,
(b) the noise statistics through \done{$c_s$}, (c) the choice of $\bA$
through $\mu(\bA)$, (d) the depth of each leaf through $\bits{L}$, (e) the
size of each leaf through $\sum_{i,j \in L}\anglebr{\bA^{(i)},\bA^{(j)}}$ and
(f) the parameter $\delta$. Ideally we would like to minimize $\RiskN(\S)$ to
obtain \wub{$\tSN$} in \eref{eqn:idealEst}. Since $\RiskN(\S)$ is bounded by
$\hRiskN(\S) + \penN(\S)$, minimizing the bound \eref{eqn:finalRiskBound}
will ensure \kk{that our estimate $\hSN$ in \eref{eqn:Shat} is as close to
$\tSN$ in \eref{eqn:idealEst} as possible.}
{In order to minimize the risk difference in
\eref{eqn:finalRiskBound}, one needs to choose an estimate $\S \in \sSM$ that
has the least $\penN(\S)$ in \eref{eqn:Phi}. The penalty term in \eref{eqn:Phi}
is directly proportional to the number of leaves in the partition $\pi(\S)$
and the size of each leaf through the term $\sum_{i,j \in
L}\anglebr{\bA^{(i)},\bA^{(j)}}$. } As a result, searching for an estimate
\wub{$\S \in \sSM$} that minimizes \kk{\eref{eqn:Shat}}
will favor estimates with few, deep leaves that hone in on the boundary of
the level set.

\wub{The} theoretical analysis of our method is significantly different from
the analysis in \cite{willett:levelset} because of the \wub{statistics of the
noise term $\bn'$ in our problem. However, this} only changes the way the
penalty is defined in our setup. As a result, we can adapt the computational
techniques discussed in \cite{willett:levelset} to compute our estimator in
an efficient way. Our method is computationally efficient since the proxy
computation needs at most $O(KN)$ operations (fewer if $\bA$ \wub{is
structured}; e.g., $\bA$ is a Toeplitz matrix) and the level set estimation
method needs $O(N \log N)$ operations\wub{, as noted in
\cite{willett:levelset}}.

\subsection{Performance analysis}
\done{As discussed earlier in Sec.~\ref{Sec:ProbForm}, our eventual goal is
to estimate the continuous-domain level set $\Sstar$ from discrete
measurements $\by$. In this section, we show that estimating the
discrete-domain level set $\SstarN$ helps us achieve this goal by
establishing that ($i$) $\SstarN \longrightarrow \Sstar$ as $N
\longrightarrow \infty$ and ($ii$) providing conditions, as a function of
problem parameters, under which the discrete-domain level set estimate
obtained according to \eref{eqn:Shat} approaches $\Sstar$.} {\wub{To this
end, we} can \wub{utilize} the results of Theorem~\ref{thm:CSLevelSets} to
upper bound the expected excess risk $\expectbr{R(\hSN)-R(\Sstar)}$\wub{,}
taken with respect to the \wub{noise distribution}, \wub{in terms of the}
problem parameters, where
\begin{align*}
R(S) = \int_{\squarebr{0,1}^d} \paranbr{\gamma-f(x)} \squarebr{\ind{x \in S} - \ind{x \notin S}} dx
\end{align*}
is the definition of risk in continuous-domain.
\wub{The expected excess risk is a measure of the effectiveness
of our level set estimator. Before studying it,} we make certain assumptions
about the smoothness of $f$ in the vicinity of the level set boundary.
%
%
{Let \kk{$\dou \Sstar$} represent the level set boundary
corresponding to \kk{$\Sstar = \setbr{x: f(x) > \gamma}$}}. We \wub{assume} that
$f$ is in a box-counting function class $\Dbox(\kappa,\gamma,c_1,c_2)$
for $c_1,c_2 > 0$ and $1 \leq \kappa \leq \infty$
\cite{willett:levelset} such that the following hold:
\begin{enumerate}[(a)]
\item If we partition $[0,1]^d$ to $m^d$ equisized
cells for $m \leq M$, \wub{with} each of them \wub{having} a sidelength of $1/m$ and volume
$m^{-d}$, then the number of such cells intersected by the level set boundary
\kk{$N_{\Sstar}(m) \leq c_1 m^{d-1}$}. This ensures that \kk{$\dou\Sstar$} varies smoothly and is not an
irregular, space-filling curve.
\item \done{For all dyadic $m$, let
\begin{align}
\Smstar = \argmin_{\S \in \sS_m} \lambda\paranbr{\Delta\paranbr{\S,\Sstar}} \label{eqn:Smstar}
\end{align}
be a candidate in $\sS_m$ that minimizes the symmetric difference between any $S \in \sS_m$ and the true level set $\Sstar$ in terms of the Lebesgue measure $\lambda$.}
For this $S_m^*$, the excess risk in \wub{the}
continuous domain follows
\kk{
\begin{align}
\err\paranbr{\wub{\Smstar},\Sstar} = R(\Smstar)-R(\Sstar) \deq \int_{\Delta(\Sstar\!,\Smstar)} \abs{\gamma-f(x)} dx\leq c_2m^{-\kappa}. \label{eqn:approxError}
\end{align}
}
\end{enumerate}
\kk{Parameter $\kappa$ and the assumption on the excess risk in
\eref{eqn:approxError} allow us to study the fluctuations of $f$ around
\kk{$\dou\Sstar$} and thus examine the behavior of
$\err\paranbr{\wub{\Smstar},\kk{\Sstar}}$ in the vicinity of the level set boundary.
If $f$ exhibits a very small fluctuation around \kk{$\dou\Sstar$}, then
$\err\paranbr{\wub{\Smstar},\kk{\Sstar}}$ is small even if the symmetric difference
$\Delta(\kk{\Sstar\!},\Smstar)$ is very large, since the excess risk is weighted by
how close is $f$ to $\gamma$. In other words, a high value of $\kappa$
indicates that $f$ varies very smoothly around the level set boundary and a
low value of $\kappa$ means that there is a jump in $f$ around \kk{$\dou\Sstar$}.
}

\kk{Recall that $\SstarN$ is obtained by partitioning the space $[0,1]^d$ to $N$ equisized cells of sidelengths $N^{-d}$ and assigning each cell to be inside or outside of the level set. From
\eref{eqn:approxError}, for $m = N^{1/d}$,
\begin{align*}
\Riskc(\SstarN) - \Riskc(\Sstar) \leq c_2 N^{-\kappa/d}.
\end{align*}
Thus $\SstarN \longrightarrow \Sstar$ as $N \longrightarrow \infty$ and estimation of $\Sstar$ via $\SstarN$ is reasonable.
}

\kk{To achieve the results of Theorem~\ref{thm:expectedExcessRiskBounds}
\done{stated below},
we adapt the prefix code proposed in \cite{willett:levelset,scott2006minimax}. According to \cite{willett:levelset,scott2006minimax}, a leaf $L$ of a level set at depth $j$ of the tree can be uniquely encoded using a total of $j(\log_2 d + 2)+1$ bits. Specifically, one needs $j+1$ bits to encode the depth of the leaf, $j$ bits to encode whether each of its ancestors corresponded to a left or a right branch of the tree, and $j \log_2d$ bits to encode the orientation of each of the $j$ branches.}

{Before we state our main theorem, let us clarify
the notation used in \wub{the following}. For a given set of sequences $a_n$
and $b_n$, $a_n \preccurlyeq b_n$ implies that there exists a constant $C>0$
such that $a_n \leq Cb_n$ for all $n$ and \kk{$a_n \asymp b_n$ implies that there exists constants $C_1$ and $C_2$ such that $C_1a_n \leq b_n \leq C_2a_n$ for all $n$}.}

\begin{theorem}[\kk{Upper bound on the expected excess risk}]
\label{thm:expectedExcessRiskBounds}
\kk{If $f \in \Dbox(\kappa,\gamma,c_1,c_2)$ is discretized according to \eref{eqn:discretization}, $-B \leq f(x) \leq
B$ for $x \in [0,1]^d$, $-B \leq \gamma \leq B$, and the estimate $\hSN$ is chosen according to \eref{eqn:Shat} with $\penN(\hSN)$ defined according to Theorem.~\ref{thm:CSLevelSets}, then, for a given $\bA$, $d \geq 2$, and for $M \succcurlyeq \paranbr{\frac{N}{\norm{\bA}_2^2\log N}}^{1/d}$,
}
\begin{align}
\expect\squarebr{\Riskc(\hSN)-\Riskc(\Sstar)}
&\preccurlyeq  \paranbr{\frac{\norm{\bA}_2^2\log N}{N}}^{\frac{\kappa}{2\kappa + d -2}} + \mu(\bA) \norm{\f}_1 \label{eqn:expectedRiskBound}
\end{align}
where the expectation is with respect to the \wub{noise distribution},
$\norm{\bA}_2$ is the spectral norm of $\bA$\wub{, $\|\bA\|_2 \deq
\sqrt{\lambda_{\max}\left(\bA^T\bA\right)}$,} and $\mu(\bA)$ is the
worst-case coherence of $\bA$.
\end{theorem}

The proof of this theorem is given in Section~\ref{Sec:PerfAnal}. \kk{This theorem tells us how the expected excess risk scales with the dimensionality $N$ of the underlying signal $\f$, the $\ell_1$ norm of $\f$, the choice of $\bA$ and the smoothness of the underlying function around the level set boundary through the parameter $\kappa$. For a unitary matrix $\bA$, $\norm{\bA}_2 = 1$ since its \done{singular values} are all equal to $1$, $\mu(\bA) = 0$ and $\expect\squarebr{\Riskc(\hSN)-\Riskc(\Sstar)}\preccurlyeq  \paranbr{\frac{\log N}{N}}^{\frac{\kappa}{2\kappa + d -2}}$, which is the minimax optimal rate derived in \cite{willett:levelset} without the projection matrix $\bA$. Since in practice $\bA$ is dictated by the physics of the measurement system, it is not always unitary. In such cases, the above theorem tells us how any given $\bA$ increases this bound. For some $\bA$, such as the one discussed in the following section, $\mu(\bA)
\longrightarrow 0$ as $N$ and $K$ go to $\infty$ since $\bA^T\bA
\stackrel{N,K \longrightarrow \infty}{\longrightarrow} \bI$. Note that \eref{eqn:expectedRiskBound} can be specified in terms of the continuous-domain function $f$ by noting that $\norm{\f}_1 \leq N\norm{f}_{L_1}$, although it is a loose bound when the function $f$ is not completely positive (or negative).}



\begin{corollary}[Performance with random projections]
\label{cor:GaussianA}
If the entries of $\bA \in \reals^{K \times N}$ are drawn from $\Gaussian{0,1/K}$, and the columns of $\bA$ are normalized to have unit $\ell_2$ norm, then
\kk{
\begin{align}
\expect\squarebr{\Riskc(\hSN)-\Riskc(\Sstar)} \preccurlyeq  \paranbr{\frac{\log N}{N}}^{\frac{\kappa}{2\kappa + d -2}} &\squarebr{\frac{\sqrt{K}+\sqrt{N}}{\sqrt{K - \sqrt{12K \log N}}}}^{\frac{2\kappa}{2\kappa + d -2}}+ \nonumber \\
&\qquad  \frac{\sqrt{15 \log N}}{\sqrt{K}-\sqrt{12 \log N}} \norm{\f}_1  \label{eqn:expectedRiskBound_GaussianA}
\end{align}
holds with probability at least $1-\squarebr{e^{-c\paranbr{K+N}}+N^{-2}+11N^{-1}}$ as long as $60 \log N \leq K \leq \frac{N-1}{4 \log N}$.}
\end{corollary}

The proof of this corollary is provided in Sec.~\ref{proof:GaussianA}. The
result above yields an upper bound on the expected excess risk as a function
of the dimensions of the projection operator $\bA$ and \kk{$\norm{\f}_1$}.
\wub{In words, this} corollary \wub{states} that the expected excess risk
\wub{in the case of random Gaussian projections} is minimized if the number
of measurements $K$ scales linearly with $N$ and \wub{increases} if $K$
scales \wub{sublinearly} with $N$. Dependence of \wub{the estimator's}
performance on the \kk{$\ell_1$ norm} of $\f$ \wub{is due to the interference
term $\left(\bA^T\bA-\bI \right)\f$ that arises during the} proxy
construction. \wub{The foregoing results provide} key insights into ways by
which we can minimize the expected excess risk and improve performance, as
discussed in detail in the following section.

\done{We conclude our discussion of
Theorem~\ref{thm:expectedExcessRiskBounds} by pointing out that practically
meaningful lower bounds for this problem are unknown at this time, but would
be the subject of a future investigation. In addition, note that our focus in
here has been on very fast, easily implementable methods for real-time
estimation of level sets. While significantly slower methods could
conceivably be developed to potentially provide lower errors, such methods
would not be able to compete with our proposed approach in terms of the
computational costs (see, e.g., Sec.~\ref{Sec:expts}).
}


\section{Performance improvement via projected median subtraction}
\label{Sec:DCOffsetSubt}
So far we \wub{have shown} that the signal-dependent
interference term in \eref{eqn:proxy} leads to a penalty term proportional to
$\norm{\f}_1$ in \eref{eqn:finalRiskBound}. This implies that the
interference in $\bz$ and thus the performance of our method \kk{may worsen with the increase in $\norm{\f}_1$, which is indeed confirmed by the experimental results in Sec.~\ref{Sec:expts}.}
To find a way to
minimize the signal-dependent interference, let us write $\f = \btf +
\lambda\ones$, where $\lambda$ is a constant DC offset such that
\begin{align}
\norm{\btf}_1 \leq \norm{\f}_1.
\label{eqn:norm_subt}
\end{align}
If we have access to an estimate $\hlambda$ of $\lambda$, then we can
minimize the signal-dependent interference by subtracting \kk{a projection of this constant offset} to obtain
\begin{align*}
\bty &= \by - \bA{\hlambda \ones} = \bA\paranbr{\btf+\lambda\ones} + \bn - \bA\hlambda\ones\\
&=\bA\paranbr{\btf+\paranbr{\lambda-\hlambda}\ones} + \bn \approx \bA\btf + \bn,
\end{align*}
assuming that $\hlambda \approx \lambda$. The proxy observations in this case reduce to
\begin{align*}
\btz &= \bA^T\bty \approx \bA^T\bA\btf + \bA^T\bA\bn = \btf + \paranbr{\bA^T\bA-\bI}\btf + \bA^T\bA\bn.
\end{align*}
Since $\SstarN = \{i: f_i > \gamma\} \deq \{i: \tf_i > \tgamma\}$, where $\tgamma = \gamma-\lambda$, we can estimate $\SstarN$ from $\btz$ using our level set estimation method discussed in the previous section.

\kk{If we let $\lambda$ to be
the median of $\f$, then we can easily show that \eref{eqn:norm_subt} holds
for this particular choice of $\lambda$. Note that if $\lambda$ is the median of $\f$, then half the pixel values of $\f$ are below the median and half of the pixel values are above the median. Let $\sG = \setbr{i: f_i > \lambda}$ and $\sG^c = \setbr{i: f_i < \lambda}$. The cardinality of $\sG$ is $\abs{\sG} = N/2$ for $N$ even\footnote{\kk{We do not consider $N$ to be odd since our recursive dyadic partitions require $N$ to be in powers of two.}}. By the definition of median, $\abs{\sG} = \abs{\sG^c}$. Then
\begin{align*}
\norm{\btf}_1 &= \norm{\f - \lambda \ones}_1 = \sum_{i \in \sG} \abs{f_i - \lambda} + \sum_{i \in \sG^c} \abs{f_i - \lambda}\\
&= \sum_{i \in \sG} \paranbr{f_i - \lambda} + \sum_{i \in \sG^c} \paranbr{\lambda-f_i}
= \sum_{i \in \sG} f_i + \sum_{i \in \sG^c} -f_i - \abs{\sG} \lambda + \abs{\sG^c} \lambda\\
&= \sum_{i \in \sG} f_i + \sum_{i \in \sG^c} -f_i\\
& \leq \sum_{i \in \sG} \abs{f_i} + \sum_{i \in \sG^c} \abs{f_i} = \norm{\f}_1.
\end{align*}
}
In practice, however, estimation of
the median of $\f$ from $\by$ might be hard, though the estimation of the
mean of $\f$ might be tractable. For
instance, if we construct $\bA^\prime = \left[ \begin{smallmatrix} \ones^T\\
\bA \end{smallmatrix} \right]$ \wub{(i.e., the first row of $\bA^\prime$ is
$\ones^T$)}, then $\by^\prime = \bA^\prime\f + \bn = \left[
\begin{smallmatrix} y^\prime_1\\ \by \end{smallmatrix} \right]$, and
$\hlambda = y^\prime_1/N =
\paranbr{\sum_i f_i + n_1}/N = \lambda + n_1/N$. If the \wub{observation} noise is
negligible, or if $N$ is large, then $\hlambda \approx \lambda$ and we can
perform \kk{projected mean subtraction}, instead of \kk{a projected median subtraction}, to reduce the signal-dependent interference. While
\eref{eqn:norm_subt} does not always hold if $\lambda$ is the mean of $\f$,
simulation results in Sec.~\ref{Sec:expts} suggest that \kk{projected mean
subtraction} can result in significant improvement in performance.
\section{Proofs of theorems and corollaries}
This section presents the proofs of all the theorems and corollaries stated before.

\subsection{Proof of Theorem~\ref{thm:CSLevelSets} (Concentration of risk )}
\label{sec:CSLevelSets}
Let us begin by bounding $T_1$ and $T_2$ in \eref{eqn:T1T2} separately.
\kk{Let $\hp_L = \sum_{i
\in L}\frac{1}{N}$ be the ratio of the number of observations in leaf $L$ to the total number of observations $N$.} From the statistics of $z$, we can bound $T_1$ as
follows:
\begin{align}
& T_1
\leq \frac{1}{N} \sum_{i,j: j \neq i} \abs{f_j} \abs{\anglebr{\bA^{(i)},\bA^{(j)}}} \abs{\left[\ind{\ell(L)=1}-\ind{\ell(L)=0}\right]}\ind{i \in L}\nonumber\\
&\leq  \frac{\mu(\bA)}{N} \sum_{i \in L} \sum_{j=1:j\neq i}^N |f_j| = \frac{\mu(\bA)}{N} \sum_{i \in L} \left(\sum_{j=1}^{N} \abs{f_j}-\abs{f_i}\right)\nonumber \\
&\leq \mu(\bA) \hp_L\|\f\|_{1} - \frac{\mu(\bA)}{N}\sum_{i \in L} \abs{f_i},  \label{eqn:T1bound}
\end{align}
where the \kk{second} inequality is due to the
fact that $\abs{\left[\ind{\ell(L)=1}-\ind{\ell(L)=0}\right]} = 1$ and
$\abs{\anglebr{\bA^{(i)},\bA^{(j)}}} \leq \mu(\bA)$ for all $j \neq i$.

\done{Rewriting $T_2$ in terms of \eref{eqn:z_i} and \eref{eqn:expect_z_i} we have
\begin{align*}
T_2 &=  \frac{1}{N} \sum_{i\in L} \left(\sum_{k=1}^K a_{k,i}n_k\right)\left[\ind{\l(L)=1}-\ind{\l(L)=0}\right]
= \sum_{k=1}^K b_kn_k
\end{align*}
where $b_k = \frac{1}{N} \sum_{i\in L}
a_{k,i}\left[\ind{\l(L)=1}-\ind{\l(L)=0}\right]$. Observe that $T_2$ is a
weighted sum of $K$ independent, zero-mean, subGaussian random variables. It
then follows from a Hoeffding-like inequality for a weighted sum of
independent, zero-mean subGaussian random variables
\cite[Theorem~3.3]{rudelson2013lecture} that
\begin{align}
\prob\paranbr{\abs{\sum_{k=1}^K b_kn_k} \geq \epsilon} \leq 2\exp\paranbr{\frac{-c\epsilon^2}{c_s^2\sum_{k=1}^K  b_k^2}} \label{eqn:subgaussian_bound}
\end{align}
for $\epsilon > 0$, where $c>0$ is an absolute numerical constant. Let us now
evaluate the term $\sum_{k=1}^K  b_k^2$ in the above expression as follows:
\begin{align}
\sum_{k=1}^K b_k^2 &= \sum_{k=1}^K\paranbr{\frac{1}{N} \sum_{i\in L} a_{k,i}\left[\ind{\l(L)=1}-\ind{\l(L)=0}\right]}^2 \nonumber\\
&= \frac{1}{N^2} \sum_{k=1}^K \sum_{i \in L}a_{k,i}\left[\ind{\l(L)=1}-\ind{\l(L)=0}\right]\sum_{j \in L}a_{k,j}\left[\ind{\l(L)=1}-\ind{\l(L)=0}\right] \nonumber\\
&= \frac{1}{N^2} \sum_{k=1}^K\sum_{i \in L}\sum_{j \in L}a_{k,i}a_{k,j} = \frac{1}{N^2} \sum_{i \in L}\sum_{j \in L} \anglebr{\bA^{(i)},\bA^{(j)}} \label{eqn:bj_term}
\end{align}
where the above equation is due to the fact that $\left[\ind{\l(L)=1}-\ind{\l(L)=0}\right]^2=1$. By substituting \eref{eqn:bj_term} in \eref{eqn:subgaussian_bound} and by equating the right hand side of \eref{eqn:subgaussian_bound} to $\delta_L \in (0,1/2)$ and solving for $\epsilon$, we can show that, with probability at least $1-2\delta_L$,
\begin{align}
T_2 \leq \sqrt{\frac{\log(1/\delta_L)c_s^2 \sum_{i,j \in L}\anglebr{\bA^{(i)}\wub{,\,}\bA^{(j)}}}{2cN^2}}.  \label{eqn:T2bound}
\end{align}
}

Applying the bounds in \eref{eqn:T1bound} and \eref{eqn:T2bound} to \eref{eqn:T1T2} we can see that with probability at least $1-2\delta_L$, the following holds:
\begin{align*}
\abs{\RiskN(L) - \hRiskN(L)} \leq& \left(\mu(\bA) \hp_L\|\f\|_{1} - \frac{\mu(\bA)}{N}\sum_{i \in L} \abs{f_i}\right) \nonumber \\
&+ \sqrt{\frac{\log(1/\delta_L)c_s^2 \sum_{i,j \in L}\anglebr{\bA^{(i)}\wub{,\,}\bA^{(j)}}}{2N^2}}.
\end{align*}
Thus for a given $\S \in \sSM$, the risk difference $\abs{\RiskN(\S)-\hRiskN(\S)}$ is upper bounded by summing the bound corresponding to each leaf separately. Since $\sum_{L\in \pi(\S)} \hp_L = 1$ and $\sum_{L \in \pi(\S)} \sum_{i \in L} \abs{f_i} = \norm{\f}_1$ we have
\begin{align*}
\abs{\RiskN(\S)-\hRiskN(\S)} \leq &\mu(\bA)\left(\frac{N-1}{N}\right) \norm{\f}_1 + \sum_{L \in \pi(\S)}\sqrt{\frac{\log(1/\delta_L)c_s^2 \sum_{i,j \in L}\anglebr{\bA^{(i)}\wub{,\,}\bA^{(j)}}}{2N^2}}
\end{align*}
with high probability. If we let $\delta_L = \delta2^{-(\bits{L}+1)}$ where $\bits{L}$ is the number of bits required to uniquely encode the position of leaf $L$, then it is straightforward to follow the proof of Lemma~2 in \cite{willett:levelset} to show that the bound above holds for every $\S \in \sSM$, which leads to the result of Theorem~\ref{thm:CSLevelSets}.

\subsection{Proof of Theorem~\ref{thm:expectedExcessRiskBounds} (Performance analysis)}
\label{Sec:PerfAnal}
In order to analyze the performance of our estimator, we will draw upon the proof techniques and the associated performance analyses in previous works on classification and level set estimation \cite{scott2006minimax,willett:levelset}. Note that some of the steps in our analysis that are adapted from \cite{scott2006minimax,willett:levelset} are repeated here for readability.

\kk{The proof of this theorem follows by relating the continuous-domain risk of a level set $\S \in \sSM$ to its discrete counterpart and exploiting the results from Theorem~\ref{thm:CSLevelSets}.
By expanding $\Riskc(S)$ for any $S \in \sSM$ in terms of the discretization of $f$ in \eref{eqn:discretization}, we have,
\begin{align*}
\Riskc(S) &= \int_x (\gamma-f(x)) \squarebr{\ind{x \in S}-\ind{x \notin S}} dx \\
&= \sum_{i=1}^N \int_{C_i} (\gamma-f(x)) \squarebr{\ind{C_i \in S}-\ind{C_i \notin S}} dx\\
&= \sum_{i=1}^N \paranbr{\gamma \vol{C_i} - \vol{C_i}f_i} \squarebr{\ind{C_i \in S}-\ind{C_i \notin S}}\\
&= \sum_{i=1}^N \paranbr{\frac{\gamma}{N} - \frac{f_i}{N}} \squarebr{\ind{i \in S}-\ind{i \notin S}}
\equiv \RiskN(S)
\end{align*}
where the second equality holds since $C_i$ is contained either in $S$ or in the complement of $S$.
Since $\hSN \in \sSM$, $\Riskc(\hSN) = \RiskN(\hSN)$. Let us consider some $\SN' \in \sSM$ that minimizes the penalized excess risk between any $S \in \sSM$ and the true level set $\Sstar$, i.e.,
$$\SN' = \min_{S \in \sSM} \squarebr{\Riskc(S) - \Riskc(\Sstar) + 2\penN(S)}.$$
From the definitions of $\hSN$ in \eref{eqn:Shat} and $\SN'$, and the results of Theorem~\ref{thm:CSLevelSets}, the following holds with probability at least $1-2\delta$ for $\delta \in [0,1/2]$:
\begin{align}
\Riskc(\hSN)-\Riskc(\Sstar) &= \RiskN(\hSN) -\Riskc(\Sstar) \leq \min_{S \in \sSM} \squarebr{\Riskc(S) -\Riskc(\Sstar) + 2\penN(S)}. \label{eqn:riskFifthStep}
\end{align}
\done{Let $\Omega$ denote the event that \eref{eqn:finalRiskBound} from Theorem~\ref{thm:CSLevelSets} holds for all proxy observations $\bz$.}
Since $-B \leq f(x) \leq
B$ for all $x \in [0,1]^d$ and $-B \leq \gamma \leq B$, for $\delta = 1/N$
\begin{align}
\expect&\squarebr{\Riskc(\hSN)-\Riskc(\Sstar)} = \expect\squarebr{\RiskN(\hSN)-\Riskc(\Sstar)}\nonumber \\
&= \expect\squarebr{\RiskN(\hSN)-\Riskc(\Sstar)|\Omega}\prob(\Omega) + \expect\squarebr{\RiskN(\hSN)-\Riskc(\Sstar)|\Omega^c}\prob(\Omega^c) \nonumber\\
&\leq \expect\squarebr{\RiskN(\hSN)-\Riskc(\Sstar)|\Omega} + \expect\squarebr{\RiskN(\hSN)-\Riskc(\Sstar)|\Omega^c}\frac{2}{N} \nonumber\\
&\leq \min_{S \in \sSM} \left[\Riskc(S) - \Riskc(\Sstar) + 2\penN(S)\right] + 4B \times \frac{2}{N} \label{eqn:expectedRiskDiff}
\end{align}
where the first term in \eref{eqn:expectedRiskDiff} is due to \eref{eqn:riskFifthStep} and the second term is due to the boundedness assumption on $f(x)$ and $\gamma$. Specifically,
\begin{align*}
\RiskN(\hSN) - \Riskc(\Sstar) &\equiv \Riskc(\hSN) - \Riskc(\Sstar) \\
&= \int_x \paranbr{\gamma-f(x)} \squarebr{\ind{x \in \hSN} - \ind{x \notin \hSN} - \ind{x \in \Sstar} + \ind{x \notin \Sstar}} dx\\
&\leq \int_x 4B dx = 4B
\end{align*}
since $\gamma-f(x) \leq 2B$ and $\ind{x \in S} - \ind{x \notin S} \leq 1$.
Rewriting \eref{eqn:expectedRiskDiff} we have,
\begin{align}
\expect&\squarebr{\Riskc(\hSN)-\Riskc(\Sstar)} \leq \min_{S \in \sSM} \left[\Riskc(S) - \Riskc(\Sstar) + 2\penN(S)\right] + \frac{8}{N} \label{eqn:expectER_firstEqn}\\
&\leq \min_{1 \leq m \leq M} \min_{S \in \sS_m} \left[\Riskc(S) - \Riskc(\Sstar) + 2\penN(S)\right] + \frac{8}{N} \label{eqn:expectER_secondEqn}\\
&\leq \min_{1 \leq m \leq M} \Riskc(\Smstar) - \Riskc(\Sstar) + 2\penN(\Smstar) + \frac{8}{N} \label{eqn:expectER_thirdEqn}\\
&\leq \min_{1 \leq m \leq M} m^{-\kappa} + 2\penN(\Smstar) + \frac{8}{N} \label{eqn:expectER_fourthEqn}
\end{align}
where $\Smstar$ in \eref{eqn:expectER_thirdEqn} is defined in \eref{eqn:Smstar} and \eref{eqn:expectER_fourthEqn} is due to \eref{eqn:approxError}.
}

Let us now bound $\penN(\Smstar)$ given in \eref{eqn:Phi}. To this end, let us rewrite
\begin{align}
\penN(\Smstar) = \left(\frac{N-1}{N}\right)\mu(\bA) \|\f\|_{1} + \penN'(\Smstar) \label{eqn:PhiRewritten}
\end{align}
where
\begin{align*}
\penN'(\Smstar) &= \sum_{L \in \pi(\Smstar)}\frac{1}{N}\sqrt{\frac{\left[\log(2N)+\bits{L}\log 2\right] \abs{c_u-c_{\ell}}^2\sum_{i,j \in L}\anglebr{\bA^{(i)},\bA^{(j)}}}{2}}
\end{align*}
and bound $\penN'(\Smstar)$.

To keep the \wub{notation simple}, let $\abs{L} = \wub{\left(\sum_{i \in
L}1\right)}$ be the number of pixels in leaf $L$ and $\btA_L$ be a $K \times
\abs{L}$ matrix formed by collecting the columns of $\bA$ corresponding to
the indices $i \in L$. Note that $\abs{L} = \wub{\sum_{i \in L} 1} = N\hp_L$.
Let
\begin{align}
q_L = \left[\log(2N)+\bits{L}\log 2\right] \paranbr{\abs{c_u-c_{\ell}}^2/2}\hp_L. \label{eqn:qL}
\end{align}
 Using this notation, we can write
\begin{align}
&\penN'(\Smstar) = \sum_{L \in \pi(\Smstar)}\sqrt{\frac{\left[\log(2N)+\bits{L}\log 2\right] \abs{c_u-c_{\ell}}^2\squarebr{\ones^T_{\paranbr{\abs{L} \times 1}}\paranbr{\btA_L^T\btA_L}\ones_{\paranbr{\abs{L} \times 1}}}}{2N^2}} \nonumber \\
&= \sum_{L \in \pi(\Smstar)}\sqrt{\frac{q_L}{N}\squarebr{\frac{\ones^T_{\paranbr{\abs{L} \times 1}}\paranbr{\btA_L^T\btA_L}\ones_{\paranbr{\abs{L} \times 1}}}{N \hp_L}}}
= \sum_{L \in \pi(\Smstar)}\sqrt{\frac{q_L}{N}\squarebr{\frac{\norm{\btA_L\ones_{\paranbr{\abs{L} \times 1}}}_2^2}{\abs{L}}}} \nonumber\\
&= \sum_{L \in \pi(\Smstar)}\sqrt{\frac{q_L}{N}}\frac{\norm{\btA_L\ones_{\paranbr{\abs{L} \times 1}}}_2}{\sqrt{\abs{L}}} \leq \sum_{L \in \pi(\Smstar)}\sqrt{\frac{q_L}{N}}\frac{\norm{\btA_L}_2\norm{\ones_{\paranbr{\abs{L} \times 1}}}_2}{\sqrt{\abs{L}}} \label{eqn:usingSpecNorm}\\
&= \sum_{L \in \pi(\Smstar)}\sqrt{\frac{q_L}{N}}\frac{\norm{\btA_L}_2\sqrt{\abs{L}}}{\sqrt{\abs{L}}}
= \sum_{L \in \pi(\Smstar)}\sqrt{\frac{q_L}{N}}\norm{\btA_L}_2 \leq \kk{\norm{\bA}_2}\sum_{L \in \pi(\Smstar)}\sqrt{\frac{q_L}{N}} \label{eqn:PhiPrimeEqn}
\end{align}
where the inequality in \eref{eqn:usingSpecNorm} follows from the definition of the spectral norm of $\btA_L$ given below:
\begin{align*}
\norm{\btA_L}_2 &= \max_{\bx \neq \zeros} \frac{\norm{\btA_L\bx}_2}{\norm{\bx}_2}
\geq \frac{\norm{\btA_L\ones_{\paranbr{\abs{L} \times 1}}}_2}{\norm{\ones_{\paranbr{\abs{L} \times 1}}}_2}.
\end{align*}
The term $\sum_{L \in \pi(\Smstar)}\sqrt{q_L/N}$ in \eref{eqn:PhiPrimeEqn} can now
be bounded from above by using the proof techniques in
\cite{scott2006minimax,willett:levelset}.
Previous work \cite{scott2006minimax} showed that for a binary tree with $N$ leaves at its
finest level, $\bits{L} \preccurlyeq \log N$.
Note that $\hp_L = \sum_{i\in L}
\frac{1}{N} \deq \sum_{i \in L} \int_{C_i} dx \deq \int_L dx = p_L = 2^{-j(L)}$ where $j(L)$ is the depth corresponding to leaf $L$ of the tree. By
substituting these results in \eref{eqn:qL} we
have
\begin{align}
\sum_{L \in \pi(\Smstar)}\sqrt{\frac{q_L}{N}} &\preccurlyeq \sum_{L \in \pi(\Smstar)}\sqrt{\frac{\left[\log(2N)+\log N\log 2\right] \paranbr{\abs{c_u-c_{\ell}}^2/2}2^{-j(L)}}{N}} \nonumber\\
&= \sqrt{\frac{\left[\log(2N)+\log N\log 2\right] \paranbr{\abs{c_u-c_{\ell}}^2/2}}{N}}\sum_{L \in \pi(\Smstar)} 2^{-j(L)/2} \nonumber \\
&\leq \sqrt{\frac{\left[\log(2N)+\log N\log 2\right] \paranbr{\abs{c_u-c_{\ell}}^2/2}}{N}} \sum_{j=1}^{J} T_j \sqrt{2^{-j}} \nonumber \\
&\wub{\leq\,} \kk{\sqrt{\frac{\log N}{N}}}c m^{d/2-1}, \label{eqn:boundOnqLPart}
\end{align}
where \kk{$J = \log_2 N$ is the deepest level of the binary tree}, $T_j$ is the number of leaves at depth
$j$ of the tree, $c$
is a constant \wub{that} is a function of the upper and lower bounds\wub{,
$c_u$ and $c_{\ell}$, on noise,} and \eref{eqn:boundOnqLPart} follows
straightforwardly from the proof of Theorem~6 in \cite{scott2006minimax}. By
substituting \eref{eqn:boundOnqLPart} in \eref{eqn:PhiPrimeEqn}, we have the
following:
\begin{align}
\penN'(\Smstar) &\preccurlyeq m^{d/2-1}\sqrt{\frac{\log N}{N}} \norm{\bA}_2. \label{eqn:estError}
\end{align}
From \kk{\eref{eqn:expectER_fourthEqn}, \eref{eqn:PhiRewritten} and \eref{eqn:estError}, }
\kk{
\begin{align*}
\expect&\squarebr{\Riskc(\hS_N)-\Riskc(\Sstar)} \preccurlyeq \min_{1 \leq m \leq M} \setbr{m^{-\kappa} + m^{d/2-1}\sqrt{\frac{\log N}{N}}\norm{\bA}_2 + \left(\frac{N-1}{N}\right)\mu(\bA) \|\f\|_{1}+ \frac{8B}{N}}  \\
&\preccurlyeq \min_{1 \leq m \leq M} \setbr{m^{-\kappa} + m^{d/2-1}\sqrt{\frac{\log N}{N}}\norm{\bA}_2 + \mu(\bA) \|\f\|_{1}+ \frac{8B}{N}}
\end{align*}
}
We can easily show that
\kk{$m \asymp
\paranbr{\frac{N}{\norm{\bA}_2^2\log N}}^{\frac{1}{2\kappa + d -2}}
$} minimizes the expression
above. \kk{Since $1 \leq \kappa \leq \infty$, the bound on $m$ is largest for $\kappa = 1$. Exploiting this result and the fact that $m \leq M$, we have that for $M \succcurlyeq \paranbr{\frac{N}{\norm{\bA}_2^2\log N}}^{\frac{1}{d}}
$, }
\begin{align}
\expect\squarebr{\Riskc(\hS_N)-\Riskc(\Sstar)} &\preccurlyeq  \paranbr{\frac{\norm{\bA}_2^2\log N}{N}}^{\frac{\kappa}{2\kappa + d -2}} + \mu(\bA) \norm{\f}_1.
\end{align}

\subsection{Proof of Corollary~\ref{cor:GaussianA} (Performance with random projections)}
\label{proof:GaussianA}
The proof of this
corollary is obtained by bounding the spectral norm of $\bA$ and the
worst-case coherence of $\bA$ with high probability.
Let $\btA \in \reals^{K
\times N}$ be a matrix whose entries are i.i.d. draws from
$\Gaussian{0,1/K}$. Each column of $\bA$ is then simply obtained by
normalizing the columns of $\btA$, that is, $\bA^{(i)} =
\frac{\btA^{(i)}}{\norm{\btA^{(i)}}_2}$ for $i \in \setbr{1,\ldots,N}$. The
bound on $\norm{\bA}_2$ is obtained by first showing that
\begin{align}
\norm{\bA}^2_2 \leq q \norm{\btA}_2^2 \label{eqn:norm_bA_btA}
\end{align}
for some constant $q$ and then bounding $\norm{\btA}_2$ using the results from random matrix theory. In particular, \cite{Vershynin_notes} states that the spectral norm of an $K \times N$ subgaussian matrix $\bM$ is upper bounded by $\norm{\bM}_2 \leq c\paranbr{\sqrt{K}+\sqrt{N}}$ with probability $1-\exp\paranbr{-c\paranbr{K+N}}$. This result can be straightforwardly extended to show that
\begin{align}
\norm{\btA}_2^2 \leq c^2 \paranbr{\sqrt{N/K}+1}^2 \label{eqn:normBound_btA}
\end{align}
with probability $1-\exp\paranbr{-c\paranbr{K+N}}$. We show that \eref{eqn:norm_bA_btA} holds with high probability by taking the following approach:
\begin{align*}
\norm{\bA}_2^2 &= \max_{\bx: \wub{\|\bx\|_2}=1} \norm{\bA \bx}_2^2 = \max_{\bx: \norm{\bx}_2=1} \sum_i \abs{\sum_j a_{i,j}x_j}^2 = \max_{\bx: \norm{\bx}_2=1} \sum_i \abs{\sum_j \frac{\ta_{i,j}}{\norm{\btA^{(j)}}_2}x_j}^2 \\
&= \max_{\bp: \sum_{j} p_j^2 \norm{\btA^{(j)}}^2_2=1} \sum_i \abs{\sum_j \ta_{i,j} p_j}^2
\deq \max_{\wub{\bp \neq \boldsymbol{0}}} \frac{\sum_i \abs{\sum_j \ta_{i,j} p_j}^2}{\sum_{j} p_j^2 \norm{\btA^{(j)}}^2_2}
\end{align*}
where $p_j = \frac{x_j}{\norm{\btA^{(j)}}_2}$ and $\bp = \left[
\begin{smallmatrix} p_1 & p_2 & \ldots & p_N \end{smallmatrix} \right]^T$.
Following the proofs of Lemma 1 in \cite{laurent2000adaptive} and \kk{Theorem 8}
in \cite{bajwa2011two}, we can easily show that $\norm{\btA^{(j)}}_2^2 \geq
1-\frac{\sqrt{12 \log N}}{\sqrt{K}}$ with probability at least $1-N^{-3}$ for
any $j \in \setbr{1,\ldots,N}$. Applying the union bound over every possible $j \in \setbr{1,\ldots,N}$, $\norm{\btA^{(j)}}_2^2 \geq
1-\frac{\sqrt{12 \log N}}{\sqrt{K}}$ with probability at least $1-N^{-2}$. Using this result in the above equation
we have,
\begin{align}
\norm{\bA}_2^2 &\leq \max_{\bp} \frac{\sum_i \abs{\sum_j \ta_{i,j} p_j}^2}{\sum_{j} p_j^2 \paranbr{1-\frac{\sqrt{12 \log N}}{\sqrt{K}}}} \deq \frac{1}{1-\frac{\sqrt{12 \log N}}{\sqrt{K}}} \norm{\btA}_2^2 \label{eqn:normbA_intermsof_normbtA}
\end{align}
with probability exceeding $1-N^{-2}$. By substituting \eref{eqn:normBound_btA} in \eref{eqn:normbA_intermsof_normbtA}, and applying the union bound, the following holds with probability exceeding ${1-\exp\paranbr{-c\paranbr{K+N}}-N^{-2}}$:
\begin{align}
\norm{\bA}_2 &\leq c\frac{\sqrt{N/K}+1}{\sqrt{1-\frac{\sqrt{12 \log N}}{\sqrt{K}}} } = c \frac{\sqrt{K}+\sqrt{N}}{\sqrt{K-\sqrt{12 K \log N}}}. \label{eqn:boundOnnormbA}
\end{align}

The rest of the proof follows straight from \kk{Theorem 8} of \cite{bajwa2011two} which states that
\begin{align}
\mu(\bA) \leq \frac{\sqrt{15 \log N}}{\sqrt{K}-\sqrt{12 \log N}} \label{eqn:boundOnmu(A)}
\end{align}
with probability exceeding $1-11N^{-1}$ as long as $60 \log N \leq K \leq \frac{N-1}{4 \log N}$. The bound in \eref{eqn:boundOnmu(A)} together with the bound in \eref{eqn:boundOnnormbA} and the result of Theorem~\ref{thm:expectedExcessRiskBounds} yields the result of Corollary~\ref{cor:GaussianA}.


{
\section{Relationship with plug-in methods}
\label{Sec:wavelets}
The success of wavelet-based methods in estimating a piecewise smooth function from noisy measurements suggests a potential extension of such methods to the problem of level set estimation \cite{donoho_hardthr}. For instance, one  possible approach for level set estimation from projection measurements is to first estimate the underlying \kk{signal} $\f$ from proxy measurements $\bz$ using wavelet-based denoising methods and then threshold the resulting estimate at level $\gamma$. Estimating $\f$ from $\by$ through an intermediate proxy construction step is similar to the iterative hard thresholding method in compressive sensing literature with just one iteration \cite{blumensath:acha09}. While such plug-in estimation techniques using wavelet-based methods offer practical solutions to the level set estimation problem, their estimation performances are not yet understood.
}

{
The proposed multiscale, partition based set estimation method with proxy
measurements can be thought of as a combination of an iterative hard
thresholding method with just one iteration, and wavelet-based denoising
ideas. Specifically, our partition-based method is similar in spirit to the
wavelet-based denoising ideas using the unnormalized Haar wavelet transform.
Both wavelet-based methods and our method rely on the spatial homogeneity of
the underlying \kk{signal} $\f$ to perform level set estimation. The difference
between the two methods stems from the way in which the wavelet coefficients
are thresholded in each case. While the \wub{threshold in the wavelet-based
method is} chosen to minimize the mean squared error, our method thresholds
the coefficients at levels that are tailored to the level set estimation
problem. Since the proposed method shares similar ideas with wavelet-based
methods, the proof techniques presented in this \wub{paper} could potentially
be extended to wavelet-based methods in order to characterize their
estimation performances. }

Compressive sensing theory presents a variety of algorithms such as iterative
hard thresholding \cite{blumensath:acha09}, basis pursuit \cite{chen1994bp},
orthogonal matching pursuit \cite{matchingPursuit}, LASSO
\cite{tibshirani1996regression} and total-variation based methods
\cite{twist} to reliably estimate $\f$ from $\by$. One can readily use such
algorithms to first estimate $\f$ and then threshold it \kk{or use the
method in [44]} to estimate the level set. However, there are a couple of
issues in using these plug-in methods to perform level set estimation. First,
these approaches \wub{aim} to minimize the mean squared error over the entire
image. This, however, does not \wub{guarantee minimization of} errors close
to the level set boundaries, which is critical to the characterization of
level set estimation performance. Second, the iterative nature of these
algorithms make them computationally intensive and time consuming.

{
\section{Experimental results}
\label{Sec:expts} {Due to the lack of a theoretical performance comparison
between plug-in methods and our method, we present an empirical comparison of
these methods in \wub{this section} by conducting experiments on a test
image. Simulation results discussed below demonstrate that the proposed
partition-based, multiscale method using proxy observations \wub{has the
following advantages:} (a) \wub{it} is a powerful tool to perform direct
level set estimation from projection measurements, (b) \wub{it} allows us to
exploit the spatial homogeneity of the underlying function to perform set
estimation, (c) \wub{it} performs an order of magnitude better than
thresholding methods that obtain level set estimates by simply thresholding
the proxy observations at level $\gamma$, and (d) \wub{it} yields results
that are comparable to the results obtained using wavelet-based thresholding
approaches.}

In order to test the effectiveness of our
projective level set estimator,
we conduct experiments on a test image of size $128 \times 128$, shown in
Fig.~\ref{fig:trueImage}. In these experiments, we are interested in
estimating $\gamma$-level set of this test image shown in
Fig.~\ref{fig:trueLevelSet} from noisy, projection measurements of the form
$\by = \bA\f + \bn \in \reals^K$ for $K < N = 128 \times 128$, without
reconstructing $\f$ from $\by$. The entries of the projection operator in
these experiments are drawn from $\Gaussian{0,1/K}$ and \wub{the noise is
distributed as} $\bn \sim \Gaussian{\zeros,\bI}$.
}
\begin{figure}
\centering
\begin{subfigure}[True \kk{signal} $\f \in \reals^{128 \times 128}$ such that $f_i \in \squarebr{44,239}$. We measure $K = 8192$ Gaussian random projections of this image.]
{\label{fig:trueImage}
\includegraphics[width=1.5in]{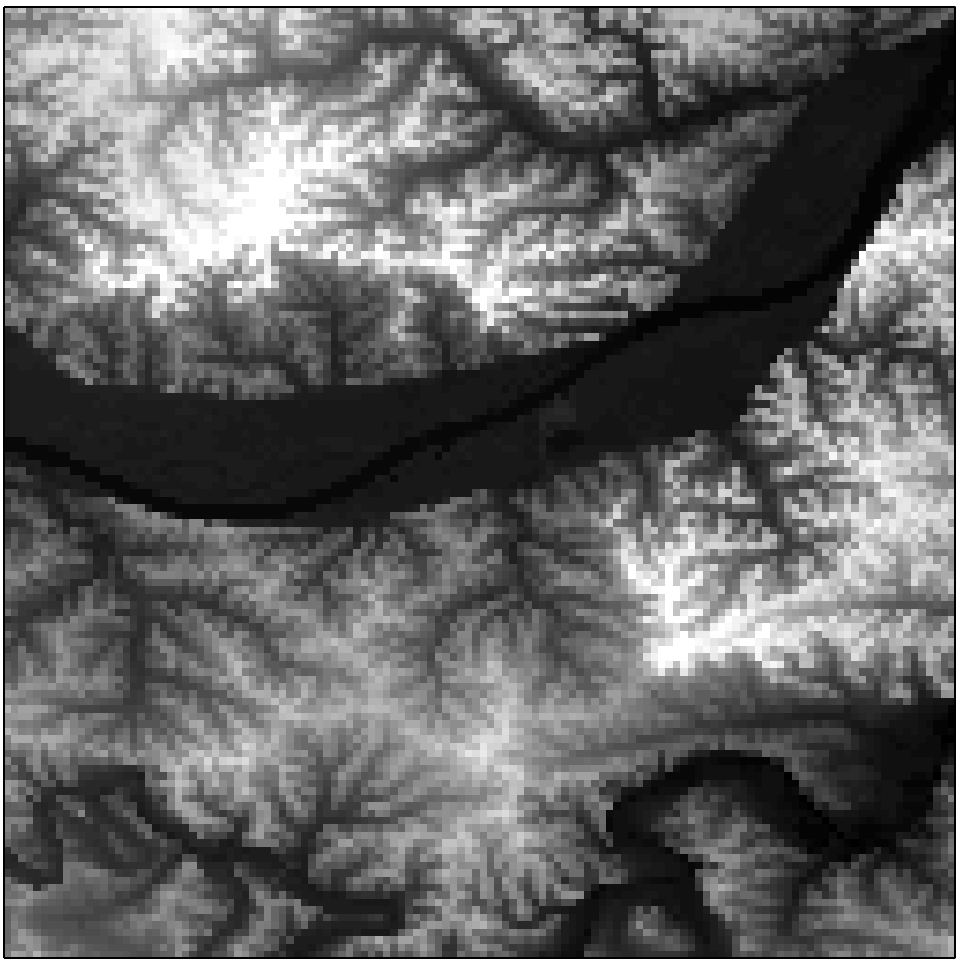} }
\end{subfigure}
\begin{subfigure}[Level set $\SstarN = \{i: f_i > 125\}$ (white pixels) such that $\abs{\SstarN} \approx 0.4285N$ where $N = 128 \times 128$.]
{\label{fig:trueLevelSet}
\includegraphics[width=1.5in]{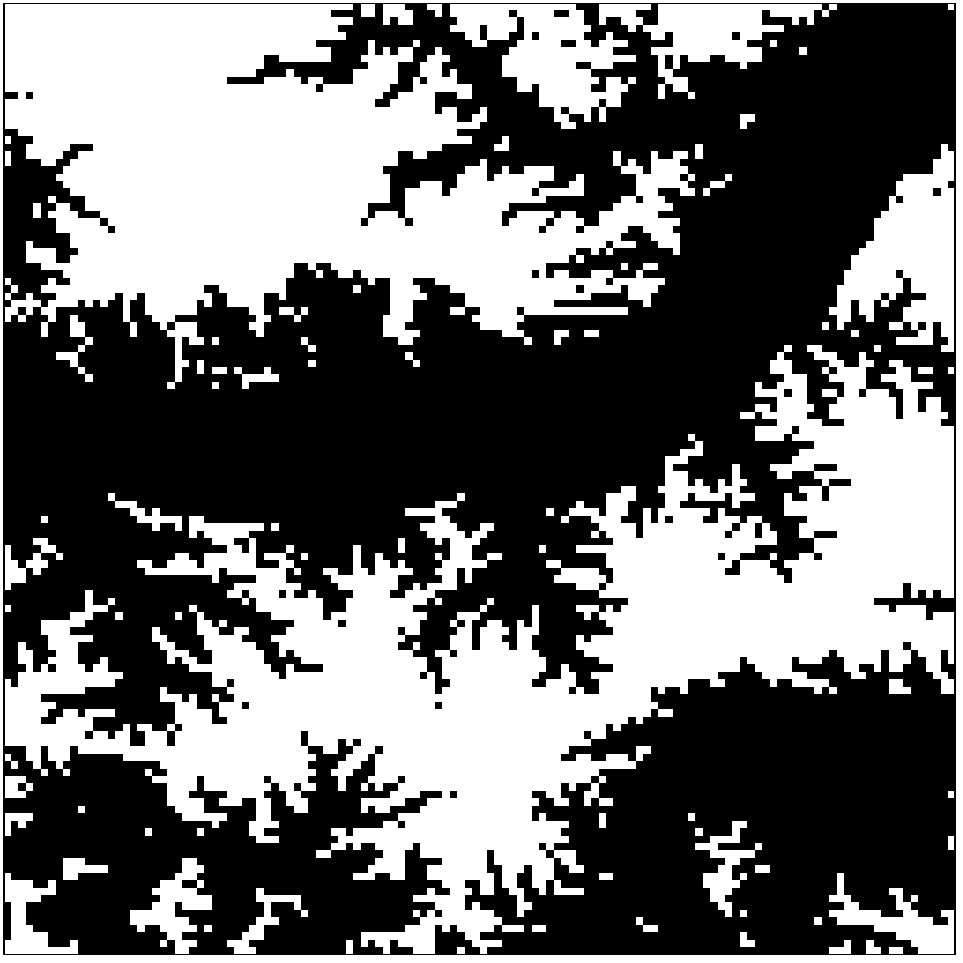}}
\end{subfigure}
\caption{Snapshots of the true signal and its \wub{desired} level set.}
\end{figure}
{We compare the performance of our method with the performances of the following approaches using the excess risk error metric defined in \eref{eqn:objective}:
\begin{enumerate}[(a)]
\item \emph{Thresholding method}, where the estimate $\hS_{\gamma}$ is
    simply obtained by thresholding the proxy observations $\bz$ at level
    $\gamma$\wub{;} that is, $\hS_{\gamma} = \{i: z_i \geq
    \gamma\}$\wub{.}
\item \emph{Risk-optimal thresholding method}, where the estimate
    $\hS_{\hgamma}$ is obtained by thresholding $\bz$ at a level
    $\hgamma$ that minimizes the excess risk\wub{;} that is,
    $\hS_{\hgamma} = \{i: z_i \geq \hgamma\}$ where $\hgamma =
    \argmin_{\gamma}\errN{\paranbr{\hS_{\gamma},\SstarN}}$\wub{.}
\item \emph{Non-iterative wavelet-based plug-in method}, where the
    estimate $\hS_w$ is obtained by first estimating $\f$ from $\bz$
    using translation invariant wavelet denoising, and then thresholding
    the resulting estimate $\bhf$ at level $\gamma$\wub{;} that is,
    $\hS_w = \{i: \hf_i \geq \gamma\}$. In these experiments we perform
    wavelet denoising using Daubechies-4 wavelets and soft thresholding,
    where the threshold is chosen to minimize the excess risk.
\item \emph{Total-variation (TV) based plug-in method}, where the
    estimate $\hS_{TV}$ is obtained according to
    $\hS_{TV} = \setbr{i:
    \hf^{(TV)}_i \geq \gamma}$. The estimate $\bhf^{(TV)}$ of the input
    image $\f$ is obtained from $\by$ by solving
\begin{align*}
\bhf^{(TV)} = \argmin_{\btf} \norm{\by - \bA\btf}_2^2 + \tau \norm{\btf}_{TV}
\end{align*}
where $\norm{\btf}_{TV}$ is the total-variation norm of $\btf$, and
$\tau$ is \wub{a} user-defined parameter that balances the log-likelihood
term and the regularization term. Algorithms such as the two-step
iterative shrinkage and thresholding (TwIST) method provide a way to
\wub{efficiently} solve for the above optimization problem \cite{twist}.
In our experiments, $\tau$ is chosen to minimize the excess risk.
\end{enumerate}
In these simulation experiments, we compute the excess risk clairvoyantly
based on the knowledge of $\f$. \kk{We obtain the estimate $\hS$ using our
projective level set estimator according to
$\hSN = \argmin_{\S \in \sSM} \hR(\S)+\tau \pen(\S)$ with a scaling factor
$\tau$, which is chosen to minimize
$\err{\paranbr{\hSN,\SstarN}}$. In these experiments, we use $M = N$.}

We evaluate the performance of all the competing algorithms discussed above,
with and without \kk{projected mean subtraction} discussed in
Sec.~\ref{Sec:DCOffsetSubt}. The number of observations used in these
experiments is $K = N/2 = 8192$. Fig.~\ref{fig:proxynoMS} shows the proxy
observations obtained without mean subtraction.
Fig.~\ref{fig:thrnoMS} shows the level set estimate obtained by simply
thresholding the proxy observations at level $\gamma$ and
Fig.~\ref{fig:riskOptimalthrnoMS} shows the estimate obtained by performing
the risk-optimal thresholding method. These results demonstrate that
thresholding noisy, proxy observations results in several false positives and
misses. Though the wavelet-based plug-in method yields better results in
comparison\wub{,} as shown in Fig.~\ref{fig:waveletnoMS}, the estimate is
still severely oversmoothed and noisy. The estimate obtained using our
projective level set estimator is shown in Fig.~\ref{fig:minenoMS}. This
approach yields lower excess risk compared to the other three approaches
discussed above, preserves some of the fine details, but still performs some
oversmoothing. Fig.~\ref{fig:twistnoMS} shows the results obtained using the
TV-based plug-in method. This method yields the best results compared to the
other approaches and yields the smallest excess risk\wub{, at the expense of
first estimating the \kk{signal}}. Fig.~\ref{fig:ERVsKnoMS} plots excess risk as
a function of the number of measurements $K < N = 16384$ for all competing
methods. These plots are obtained by averaging the results obtained over 200
different noise and \wub{projection} matrix realizations.


Figs.~\ref{fig:withMSResults}(a) through \ref{fig:withMSResults}(g) show the
improvements in results obtained because of \kk{the projected mean subtraction}. The
improvements stem from the fact that the proxy measurements are less
\wub{``noisy''} after \kk{the projected mean subtraction}. \kk{This
subtraction operation} lowers the excess risk of the estimates obtained using every
method discussed above, except for the TV-based plug-in method, which
performs very well in practice irrespective of mean subtraction. TV-based
reconstruction is in general implemented using iterative algorithms where
convergence is achieved if the mean squared error between estimates obtained
in successive iterations \wub{does} not change beyond a user-specified
tolerance value. The TwIST algorithm used in our simulation study uses the
proxy measurements to initialize the iterative process and stops iterating
when convergence is achieved. As a result, only the number of iterations to
achieve a specified convergence will change depending on the quality of the
proxy observations and not the final estimate. This explains why the TV-based
results are insensitive to projected mean subtraction.

The TV-based method seems to outperform our projective level set
estimator since we evaluate the performance of these methods based solely on
the excess risk and not on the computational resources required to achieve
that excess risk. In that sense, this comparison is \wub{somewhat} unfair. A
more meaningful comparison would be to either evaluate the excess risk
obtained within some unit time, or compare the time taken by different
approaches to achieve a desired excess risk as the problem size $N$ changes.
To make the comparison fair, we ran our projective level set
estimator for different problem sizes, used \wub{$K \approx N/3$}
observations to get our estimates, recorded the excess risk obtained in each
case, and ran the TV-based plug-in method to achieve the same excess risk in
each case. In other words, instead of using the conventional convergence
strategy in TV-based reconstruction algorithm, we stop iterating if the
excess risk is less than or equal to the one obtained using our method. We
compare the computational time required for both these methods as a function
of problem size.
{Figs.~\ref{fig:timeComp}(a) and
\ref{fig:timeComp}(b) show a $512 \times 512$ image and its corresponding
level set, respectively.} \kk{Note that the image used in above experiments is a cropped version of the image in Fig.~\ref{fig:timeComp}(a).} We cropped this image in order to get images of
different sizes. In particular, we used images of size $\ell \times \ell$
where $\ell = 76,96,116,\ldots,376$. Fig.~\ref{fig:timeComp}(c)
shows the time-gap between these two methods to achieve similar excess risks,
as a function of the number of pixels in the input image. These plots show
that the computational time taken by TV-based plug-in method dramatically
increases with problem size, where as the computation time required by our
projective level set estimator increases much more gracefully with problem
size.

\done{Before concluding, it is also important to comment on the performance
of our approach in relation to that of faster plug-in methods, such as those
based on the SVD of $\bA$. As noted in Sec.~\ref{ssec:previous}, we do not
expect such methods to perform well in the underdetermined ($K < N$) setting
for reasons outlined earlier. We have also verified this intuition through
numerical experiments (not fully reported here for space reasons). Consider,
for example, estimating the level set in Fig.~\ref{fig:trueLevelSet} by
thresholding either TSVD or Tikhonov regularized solution for the case of $K
\approx N/2$. In this setting, the excess-risks obtained using TSVD and
Tikhonov regularization-based plug-in methods are $14.26$ and $14.31$
respectively, where as the excess-risk using our proposed method is $3.593$.
This rather poor performance of SVD-based plug-in methods should not be too
surprising. Such methods operate on the assumption that signals lie near a
subspace, but a union-of-subspaces model is known to be a better model for
real-world signals \cite{Lu.Do.ITSP2008}. In contrast to SVD-based
approaches, our method performs better since the family $\S_m$ over which we
search for an estimate of the level set can be construed as a
union-of-subspaces, with each subspace in the union being formed by a set of
indices corresponding to dyadic, tree-based basis functions.}

\wub{In conclusion, the} experimental results indicate that estimating the
underlying \kk{signal} using TV regularization-based plug-in methods yields more
accurate level set estimates compared to the ones obtained using our
projective level set estimator. However, the real strengths of our method are
two-fold. First, we can reliably perform \emph{real-time} level set
estimation compared to plug-in methods as shown by the time-gap \wub{versus}
problem size plot in Fig.~\ref{fig:timeComp}(c). Second, we can
use our level set estimate to discard regions where the levels of interest
are not present and design adaptive measurement schemes to hone-in on the
regions of interest. Such an adaptive measurement scheme is especially
helpful in very high-dimensional settings where the cost of collecting
measurements and performing reconstruction \wub{tends to be} extremely high.

\begin{figure}
\centering
\begin{subfigure}[Proxy observations]
{\label{fig:proxynoMS}
\includegraphics[height=1.25in]{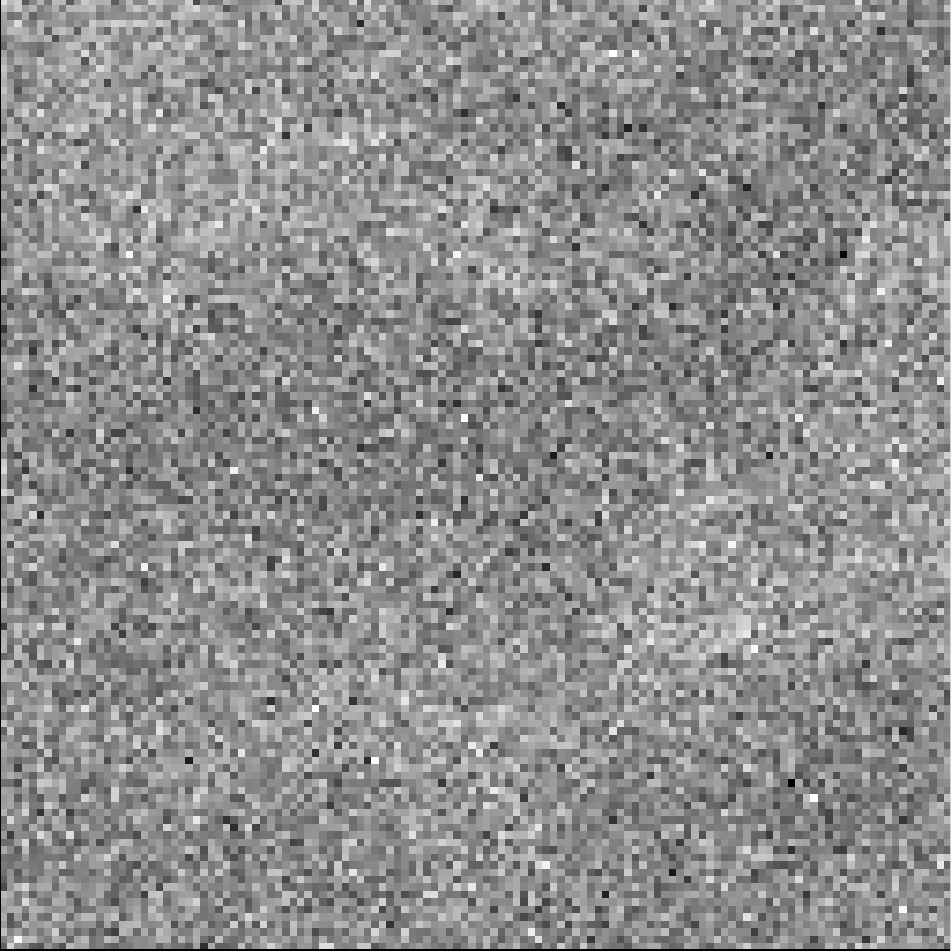}}
\end{subfigure}
\begin{subfigure}[Estimate obtained using the thresholding method; $\errN = 15.21$]
{\label{fig:thrnoMS}
\includegraphics[height=1.25in]{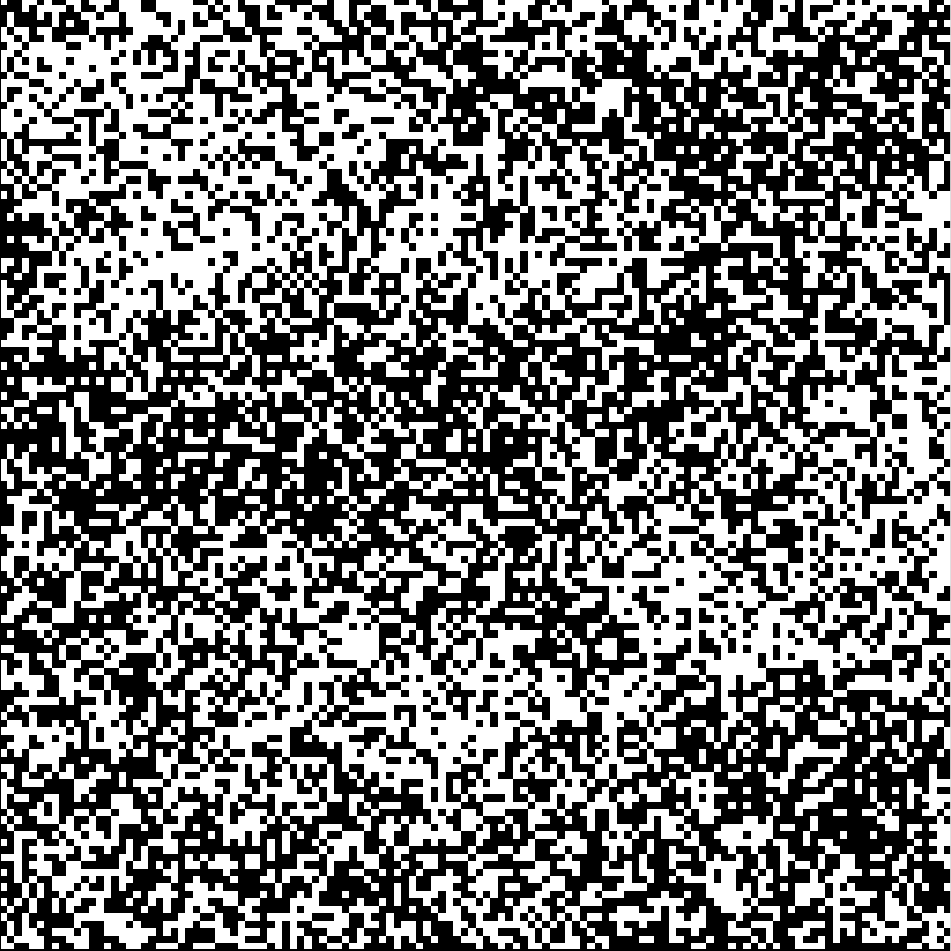} }
\end{subfigure}
\begin{subfigure}[Estimate obtained using the risk-optimal thresholding method; $\errN = 14.83$]
{\label{fig:riskOptimalthrnoMS}
\includegraphics[height=1.25in]{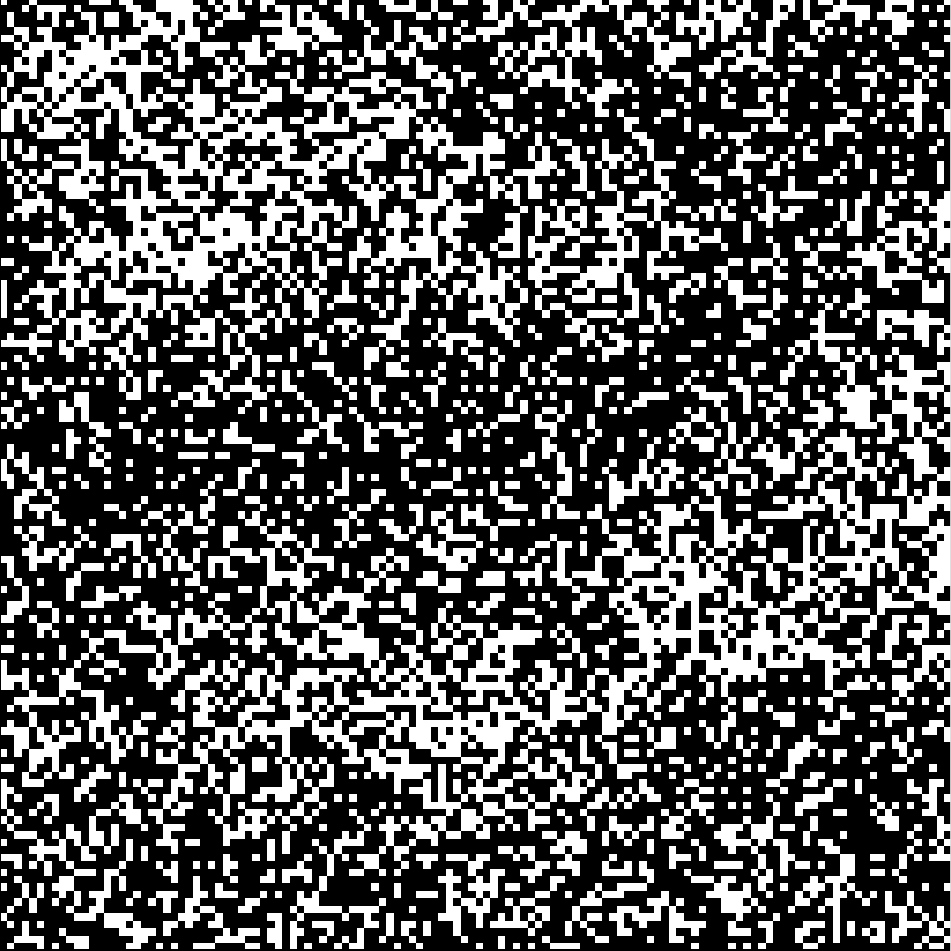} }
\end{subfigure}
\begin{subfigure}[Estimate obtained using the wavelet-based method; $\errN = 4.824$]
{\label{fig:waveletnoMS}
\includegraphics[height=1.25in]{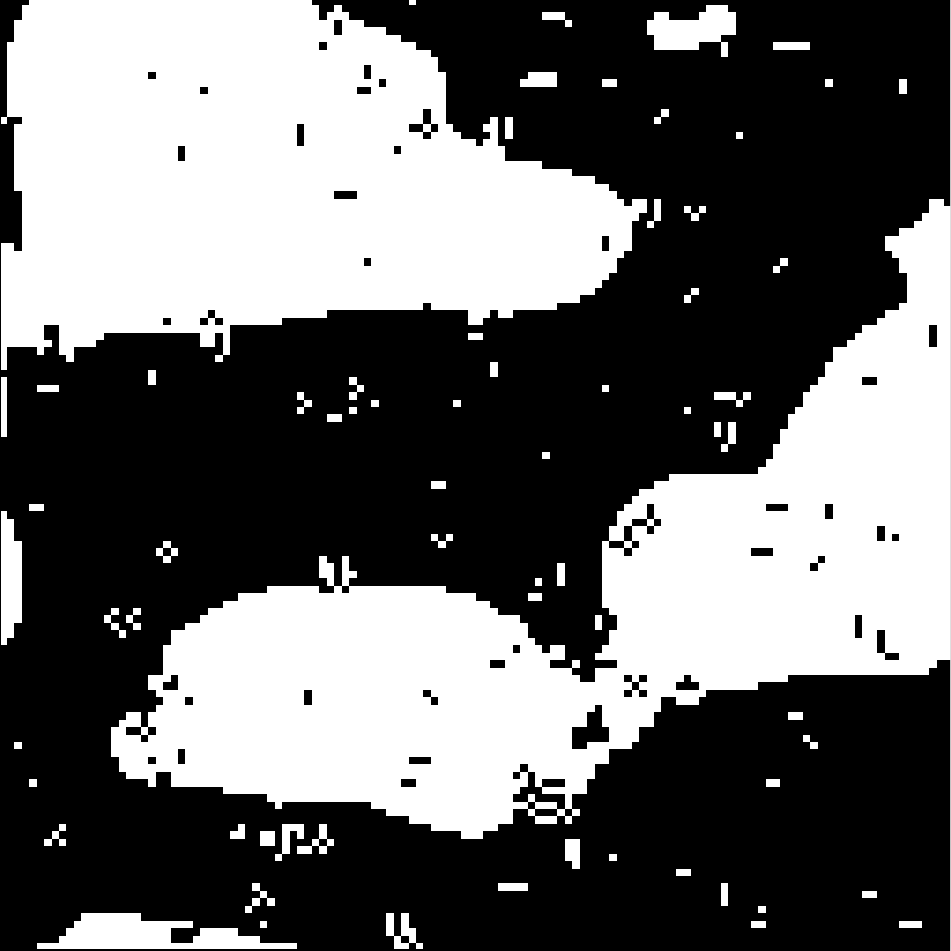} }
\end{subfigure}
\begin{subfigure}[Estimate obtained using the projective level set estimator; $\errN = 3.593$]
{\label{fig:minenoMS}
\includegraphics[height=1.25in]{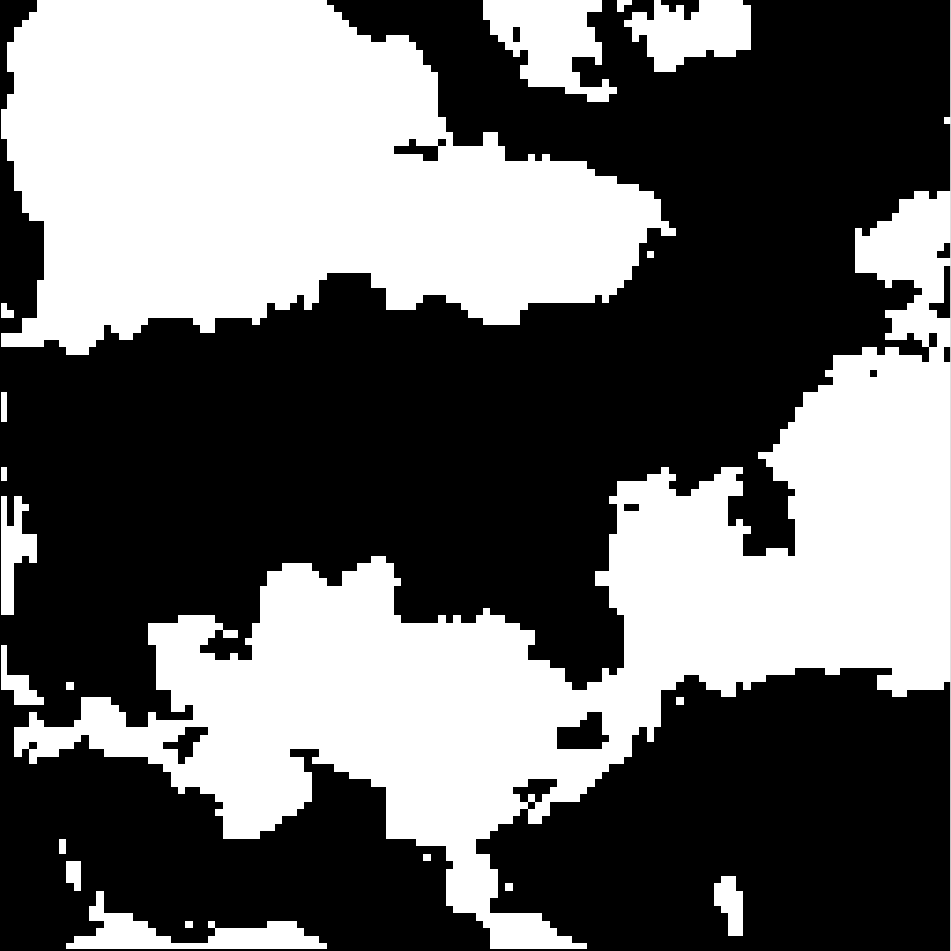} }
\end{subfigure}
\begin{subfigure}[Estimate obtained using the TV-based plug-in estimator; $\errN = 0.5596$]
{\label{fig:twistnoMS}
\includegraphics[height=1.25in]{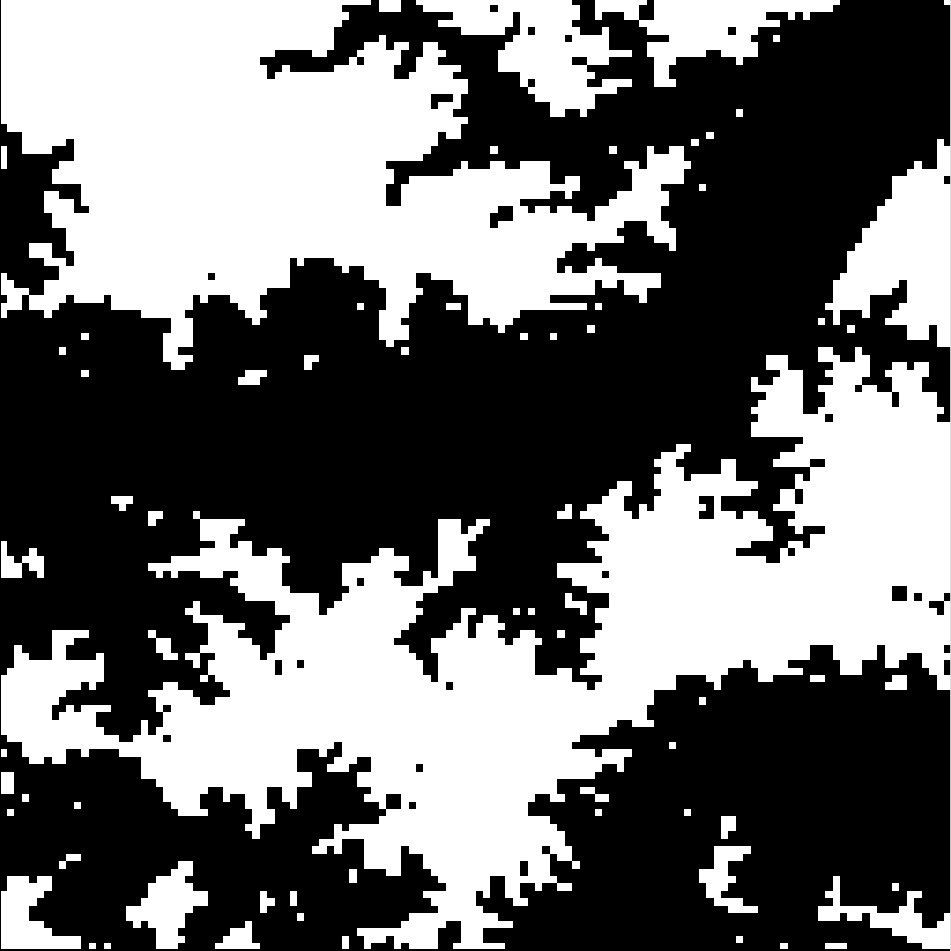} }
\end{subfigure}
\begin{subfigure}[Plot of excess risk as a function of $K < N = 16384$ without performing \kk{the projected mean subtraction}.]
{\label{fig:ERVsKnoMS}
\includegraphics[height=2.25in]{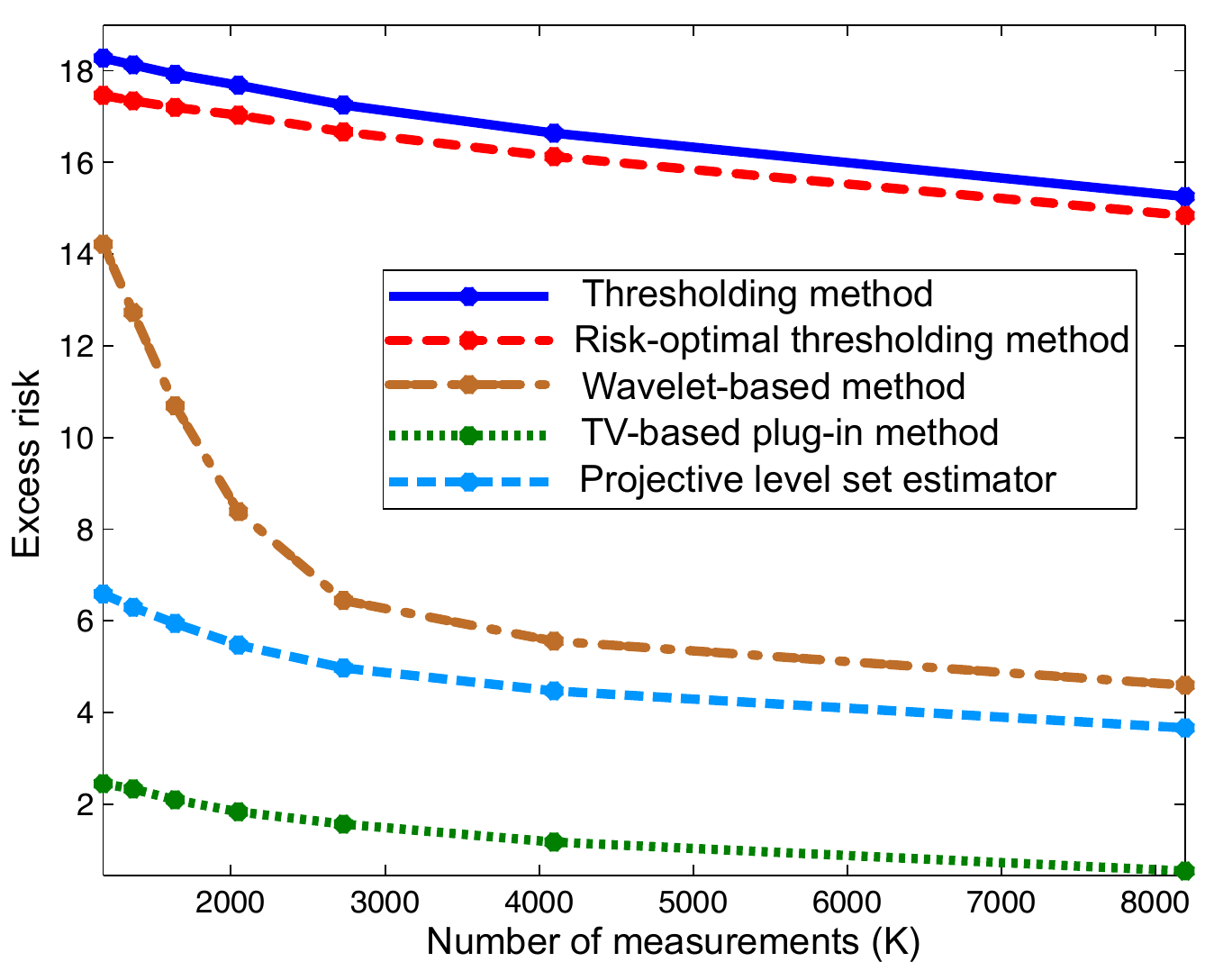} }
\end{subfigure}
\caption{\label{fig:noMSResults}Snapshots of the simulation results obtained (without performing \kk{the projected mean subtraction}) from observations of the form in \eref{eqn:probDesc}.}
\end{figure}

\begin{figure}
\centering
\begin{subfigure}[Proxy observations \wub{after \kk{projected mean subtraction}}]
{\label{fig:proxy_withMS}
\includegraphics[height=1.25in]{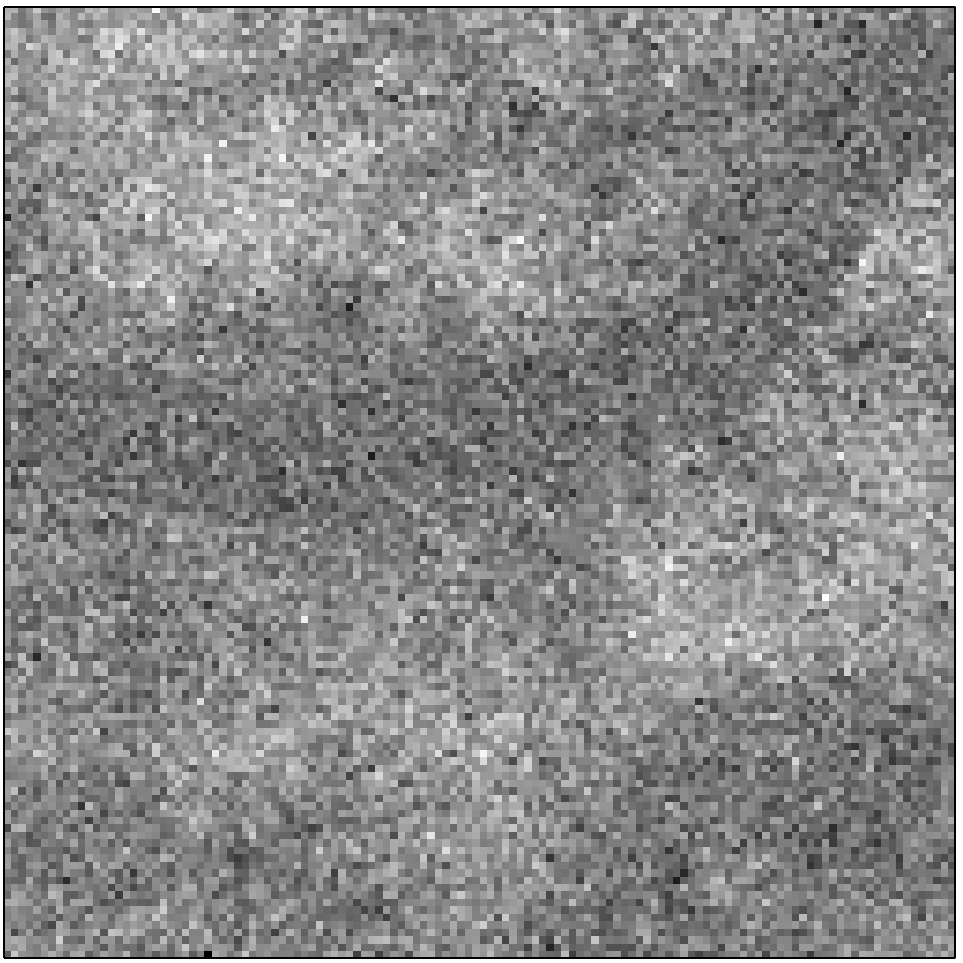}}
\end{subfigure}
\begin{subfigure}[Estimate obtained using the thresholding method; $\errN = 8.562$]
{\label{fig:thrwithMS}
\includegraphics[height=1.25in]{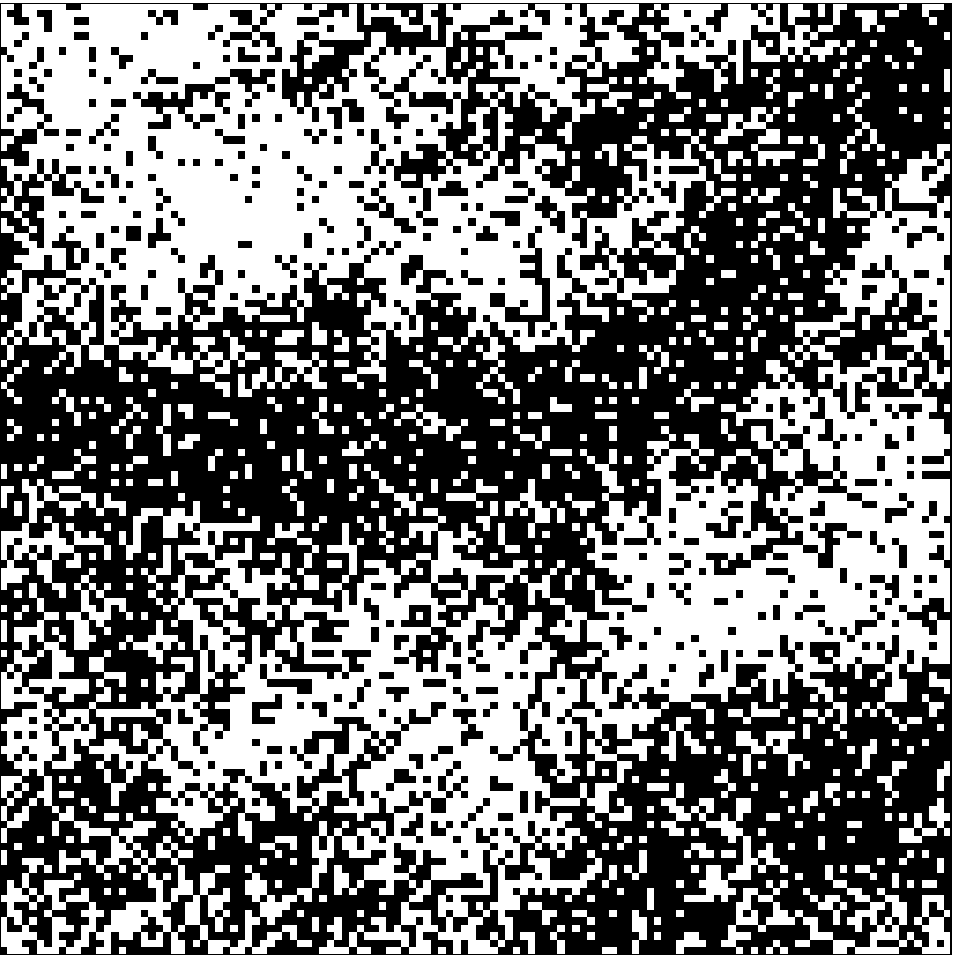} }
\end{subfigure}
\begin{subfigure}[Estimate obtained using the risk-optimal thresholding method; $\errN = 8.231$]
{\label{fig:riskOptimalthrwithMS}
\includegraphics[height=1.25in]{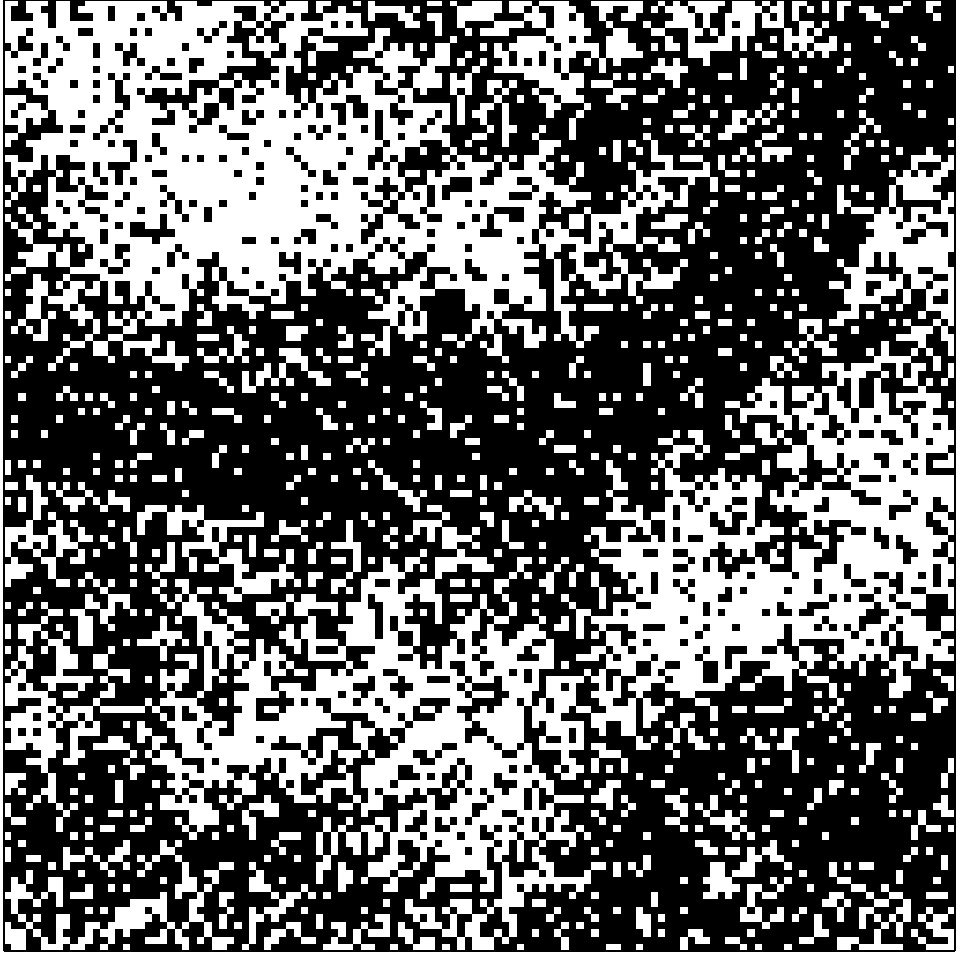} }
\end{subfigure}
\begin{subfigure}[Estimate obtained using the wavelet-based method; $\errN = 2.393$]
{\label{fig:waveletwithMS}
\includegraphics[height=1.25in]{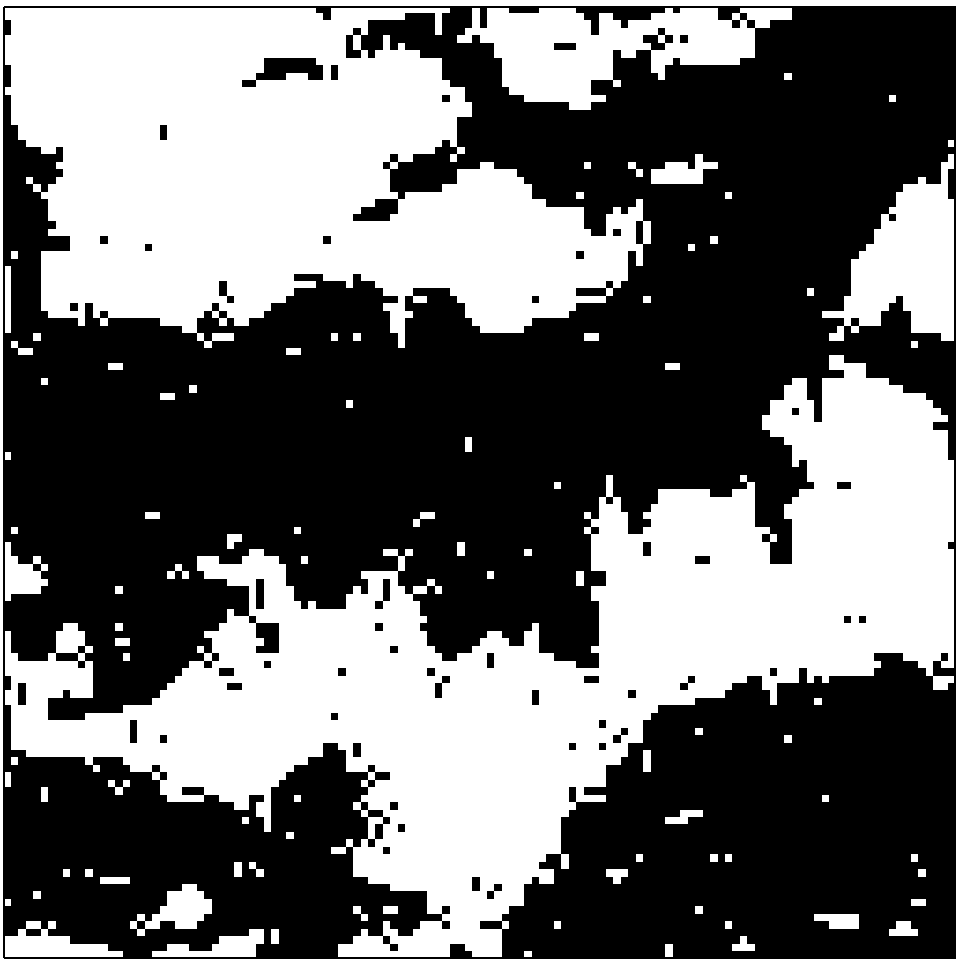} }
\end{subfigure}
\begin{subfigure}[Estimate obtained using the projective level set estimator; $\errN = 1.924$]
{\label{fig:minewithMS}
\includegraphics[height=1.25in]{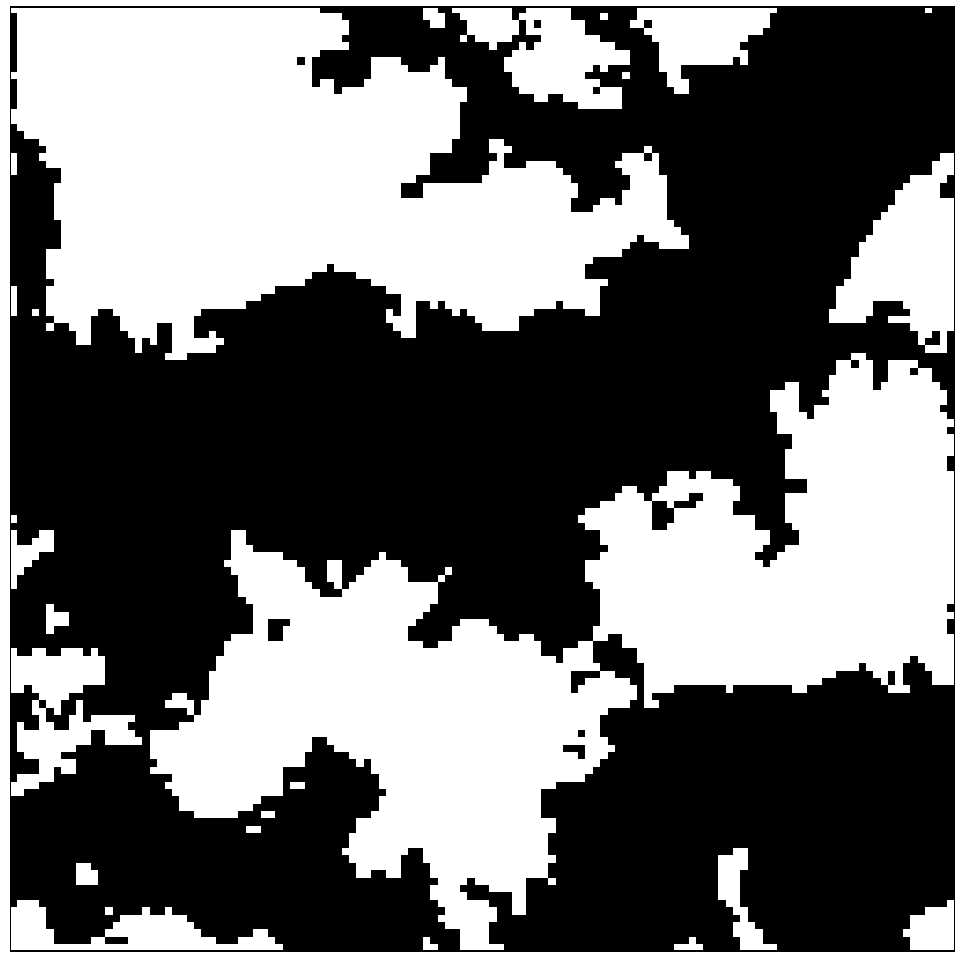} }
\end{subfigure}
\begin{subfigure}[Estimate obtained using the TV-based plug-in estimator; $\errN = 0.5593$]
{\label{fig:twistwithMS}
\includegraphics[height=1.25in]{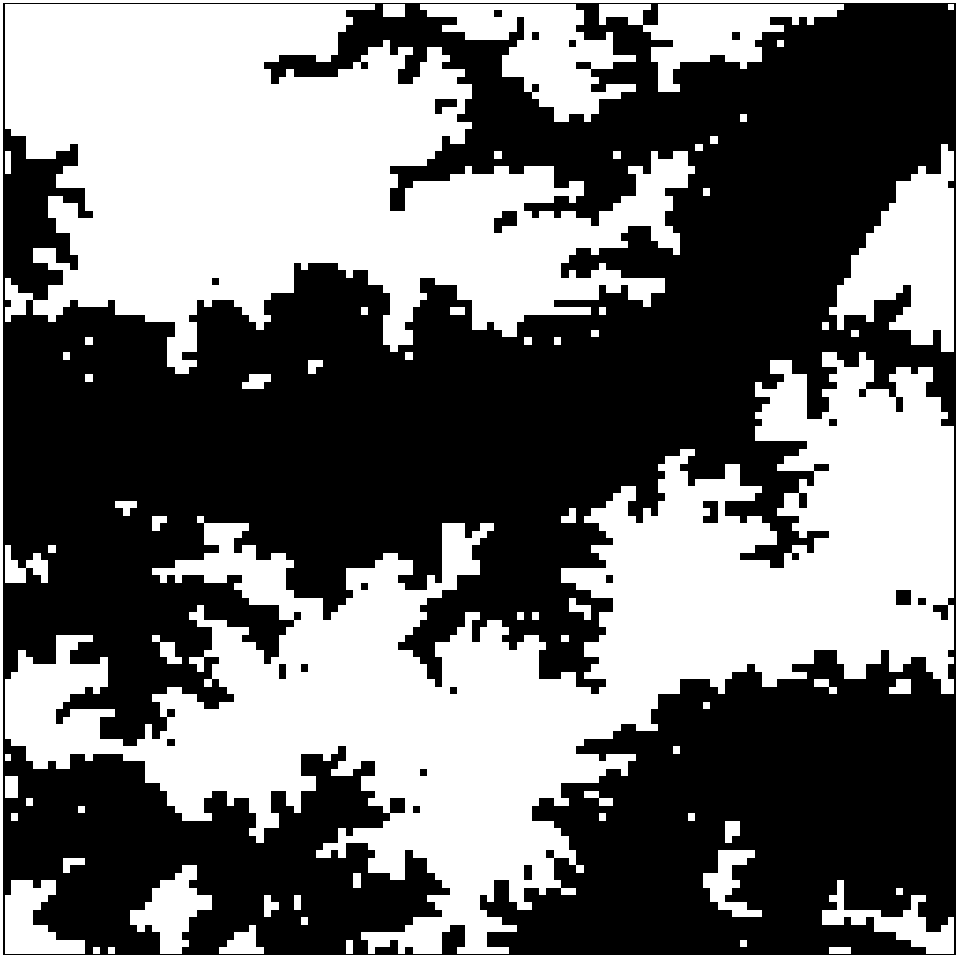} }
\end{subfigure}
\begin{subfigure}[Plot of excess risk as a function of $K < N = 16384$ after performing mean subtraction.]
{\label{fig:ERVsKwithMS}
\includegraphics[height=2.25in]{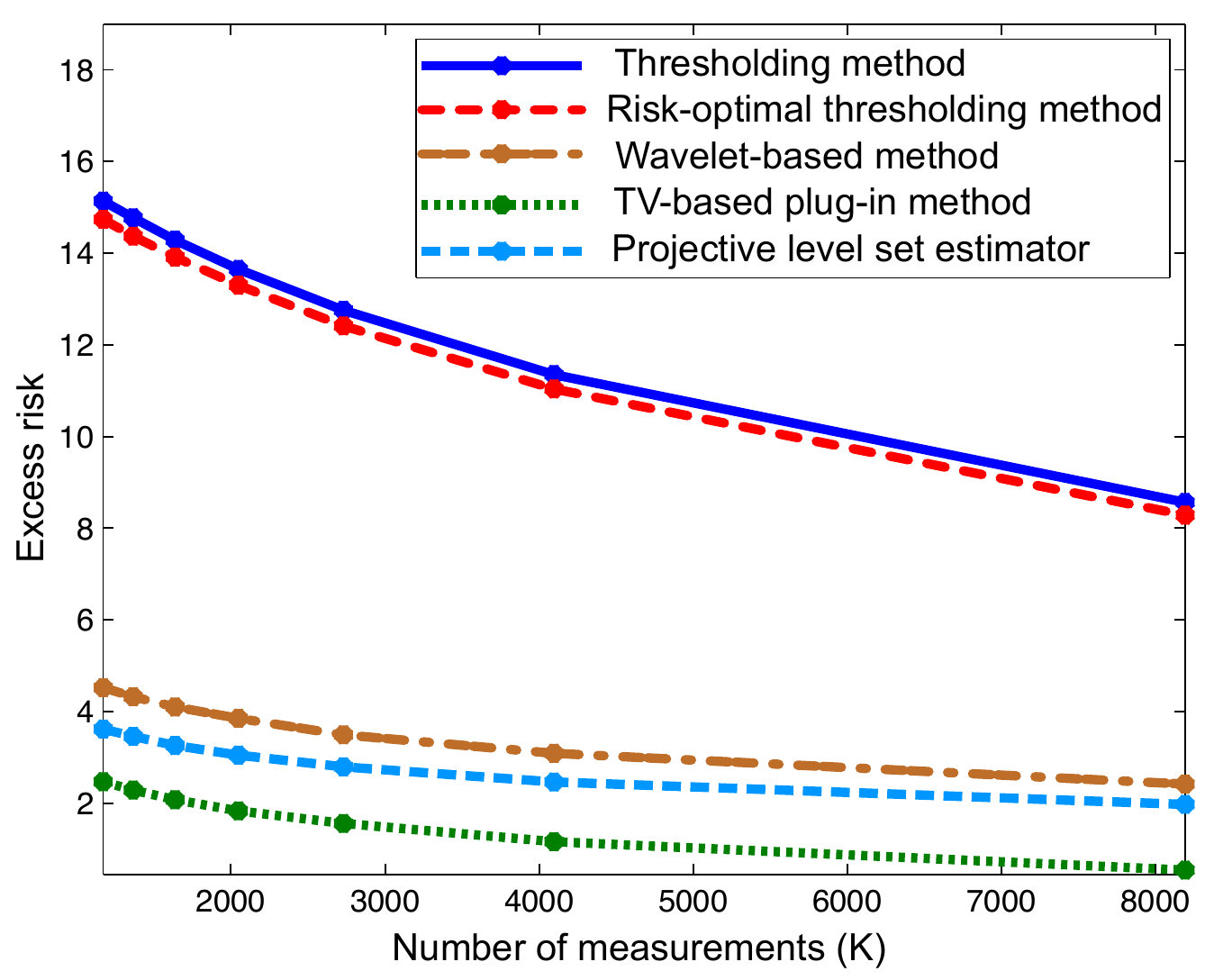} }
\end{subfigure}
\caption{Snapshots of the simulation results obtained (after performing \kk{projected mean subtraction}) from observations of the form in \eref{eqn:probDesc}.}
\label{fig:withMSResults}
\end{figure}
\begin{figure}
\centering
\begin{tabular}{ccc}
\includegraphics[width=1.5in]{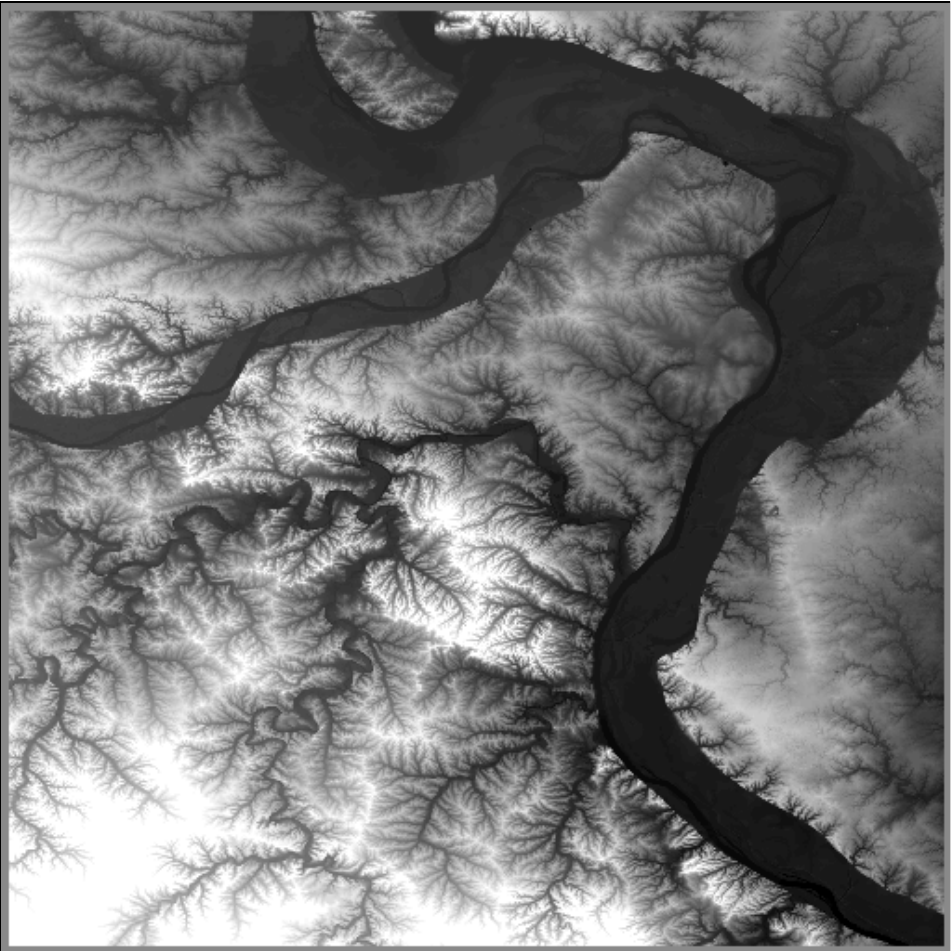}&
\includegraphics[width=1.5in]{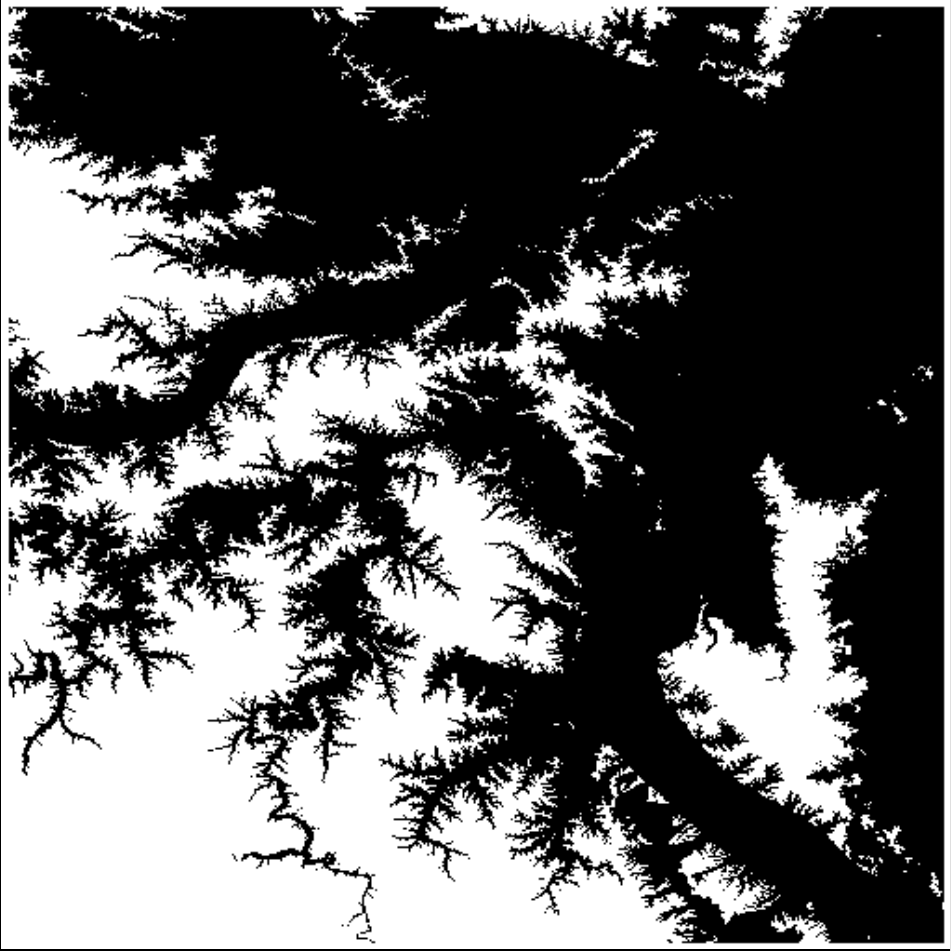}&
\includegraphics[width=1.85in]{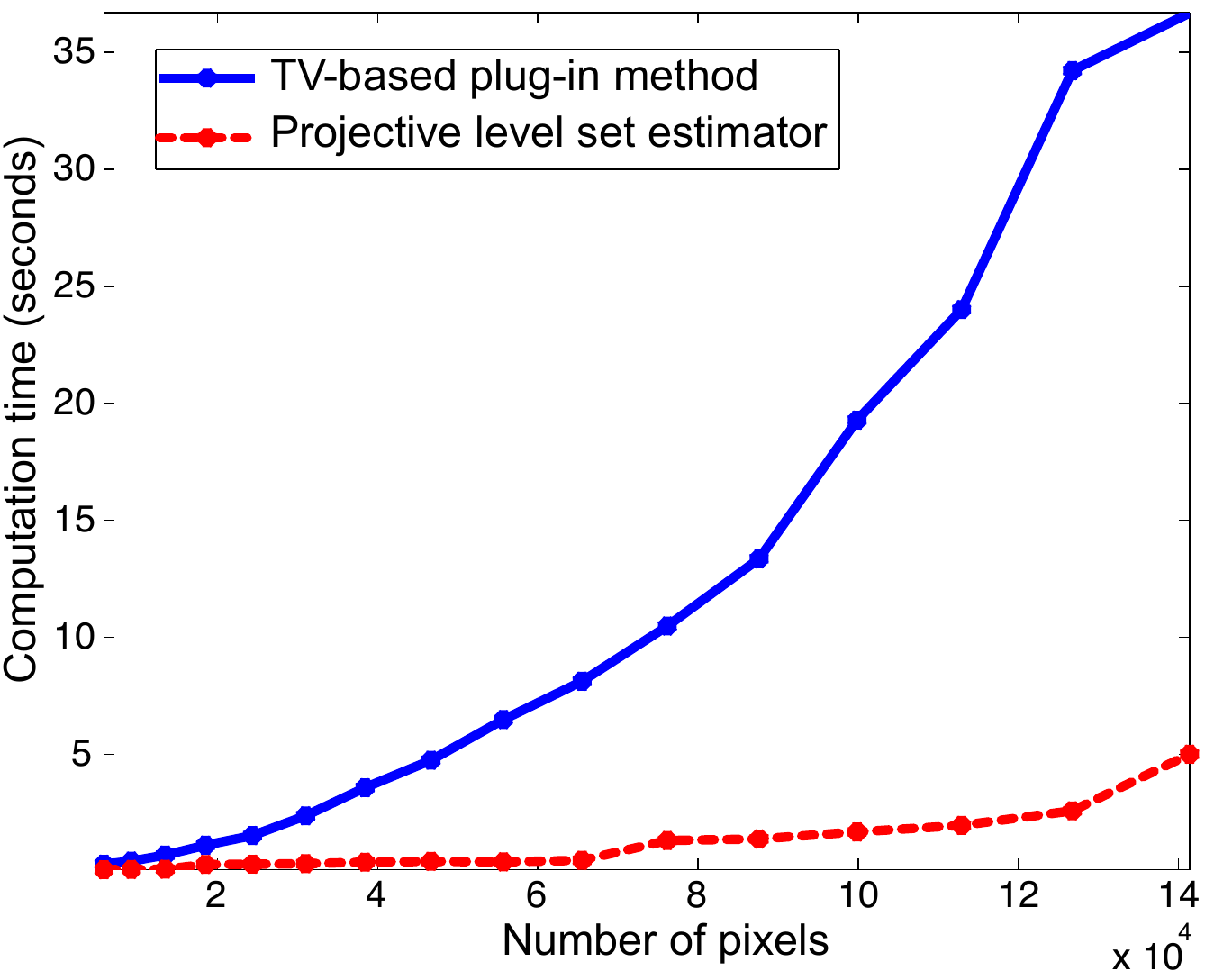}\\
(a) & (b) & (c)
\end{tabular}
\caption{\label{fig:timeComp}Comparison of the computation times of TV-based plug-in method and our projective level set estimator. (a) True \kk{signal} $\f \in \reals^{512^2}$ such that $f_i \in \squarebr{0,239}$, (b) level set $\SstarN = \{i: f_i > 125\}$, and (c) a plot of computation time as a function of problem size to approximately achieve the same excess risk. Images of different sizes ($76 \times 76$ to $376 \times 376$) cropped from (a) were used in this experiment. The plots in (c) indicate that the time it takes for the TV-based plug-in method to achieve the \wub{same} excess risk obtained by the projective level set estimator increases dramatically with problem size.}
\end{figure}
{
\section{Conclusion}
This work proposes a theoretically \wub{sound} and computationally efficient
tree-based approach for extracting level sets of a \kk{function} from projection
measurements without reconstructing the underlying \kk{function}. The simulation
results presented in Sec.~\ref{Sec:expts} suggest that the proposed method
may facilitate fast and accurate level set estimates from tomographic
projections in medical imaging, Fourier projections in interferometry, or
coded projections in compressive optical systems. One of the key advantages
of our approach is that \kk{many of the operations on the proxy data are easily parallelizable.}
\kk{For instance, in problems where the domain of the signal of interest is very large, we can compute the proxy observations, partition
the proxy data into different patches, run our estimation algorithm on each patch
separately and merge the results to identify the regions that correspond to
the level set.} In applications such as medical imaging, the time saved by
collecting fewer \wub{projection} measurements and parallelization can be
significant and crucial.}

Empirically, the accuracy of the projective level set estimate is comparable to that of a similar scheme based on wavelet thresholding or an iterative method with TV regularization. Currently, however, there is no theoretical support for these alternatives. Recent work studying the performance of so-called ``analysis regularization" \cite{vaiter2011robust,giryes2010rip} may lead to an improved understanding of theoretical performance bounds for the TV approach, but as we show here this iterative solution requires significantly more computational resources. Our approach is much more similar in spirit to the wavelet-based approach, and the theoretical techniques employed in our analysis may lead to an improved understanding of this and other fast, non-iterative approaches.
\kk{Furthermore, adaptive sampling schemes such as the one discussed in \cite{haupt2009compressive} suggest a potential extension of our method. 
Specifically, \cite{haupt2009compressive} proposes collecting noisy measurements of a sparse signal, estimating its support and collecting more measurements based on the estimated support to adaptively focus the computational resources on regions of interest. The underlying assumption in such ``distilled sensing" \cite{haupt2009distilled} schemes is sparsity. Since our level set estimation method offers a way to estimate the level set of a function without requiring sparsity, we expect it to facilitate the development of new adaptive sampling routines that perform better than the ones proposed in earlier works.}



{\bibliographystyle{siam} \sloppypar
\bibliography{CSLevelSets}}

\begin{thebibliography}{10}

\bibitem{spm:cs08}
{\em {Special Issue on Compressive Sampling}}, {IEEE} Signal Processing
  Magazine, 25 (2008).

\bibitem{ayed2005multiregion}
{\sc I.~Ayed, A.~Mitiche, and Z.~Belhadj}, {\em Multiregion level-set
  partitioning of synthetic aperture radar images}, IEEE Transactions on
  Pattern Analysis and Machine Intelligence,  (2005), pp.~793--800.

\bibitem{bajwa:jc10}
{\sc W.~U. Bajwa, R.~Calderbank, and S.~Jafarpour}, {\em Why {G}abor frames?
  {T}wo fundamental measures of coherence and their role in model selection},
  J. Commun. Netw.,  (2010), pp.~289--307.

\bibitem{bajwa2011two}
{\sc W.~U. Bajwa, R.~Calderbank, and D.~G. Mixon}, {\em Two are better than
  one: {F}undamental parameters of frame coherence}, Appl. Comput. Harmon.
  Anal., 33 (2012), pp.~58--78.

\bibitem{twist}
{\sc J.~Bioucas-Dias and M.~Figueiredo}, {\em {A New TwIST: Two-Step Iterative
  Shrinkage/Thresholding Algorithms for Image Restoration}}, IEEE Transactions
  on Image Processing, 16 (2007), pp.~2992 --3004.

\bibitem{blumensath:acha09}
{\sc T.~Blumensath and M.~Davies}, {\em Iterative hard thresholding for
  compressed sensing}, Applied and Computational Harmonic Analysis, 27 (2009),
  pp.~265--274.

\bibitem{CS:candes1}
{\sc E.~Cand\`{e}s, J.~Romberg, and T.~Tao}, {\em Robust uncertainty
  principles: Exact signal reconstruction from highly incomplete frequency
  information}, IEEE Transactions on Information Theory, 52 (2006), pp.~489 --
  509.

\bibitem{CS:candes2}
{\sc E.~Cand\'{e}s and T.~Tao}, {\em Near optimal signal recovery from random
  projections: Universal encoding strategies}, IEEE Transactions on Information
  Theory, 52 (2006), pp.~5406--5425.

\bibitem{chen1994bp}
{\sc S.~Chen and D.~Donoho}, {\em Basis pursuit}, in Signals, Systems and
  Computers, 1994. 1994 Conference Record of the Twenty-Eighth Asilomar
  Conference on, vol.~1, Oct-2 Nov 1994, pp.~41--44 vol.1.

\bibitem{cuevas2006plug}
{\sc A.~Cuevas, W.~Gonz{\'a}lez-Manteiga, and A.~Rodr{\'\i}guez-Casal}, {\em
  Plug-in estimation of general level sets}, Australian \& New Zealand Journal
  of Statistics, 48 (2006), pp.~7--19.

\bibitem{CS:donoho}
{\sc D.~Donoho}, {\em Compressed sensing}, IEEE Transactions on Information
  Theory, 52 (2006), pp.~1289--1306.

\bibitem{donoho_hardthr}
{\sc D.~Donoho and I.~Johnstone}, {\em Ideal spatial adaptation by wavelet
  shrinkage}, Biometrika, 81 (1994), pp.~425--455.

\bibitem{Duarte.Eldar.ITSP2011}
{\sc M.~F. Duarte and Y.~C. Eldar}, {\em Structured compressed sensing: {F}rom
  theory to applications}, {IEEE} Trans. Signal Processing, 59 (2011),
  pp.~4053--4085.

\bibitem{fletcher:tit09}
{\sc A.~Fletcher, S.~Rangan, and V.~Goyal}, {\em Necessary and sufficient
  conditions for sparsity pattern recovery}, {IEEE} Transactions on Information
  Theory,  (2009), pp.~5758--5772.

\bibitem{wasserman}
{\sc C.~Genovese, J.~Jin, and L.~Wasserman}, {\em Revisiting marginal
  regression}, Arxiv preprint arXiv:0911.4080,  (2009).

\bibitem{giryes2010rip}
{\sc R.~Giryes and M.~Elad}, {\em {RIP}-based near-oracle performance
  guarantees for {SP}, {CoSaMP}, and {IHT}}, {IEEE} Trans. Signal Processing,
  60 (2012), pp.~1465--1468.

\bibitem{harmany2008controlling}
{\sc Z.~Harmany, R.~Willett, A.~Singh, and R.~Nowak}, {\em {Controlling the
  error in FMRI: Hypothesis testing or set estimation?}}, in 5th IEEE
  International Symposium on Biomedical Imaging: From Nano to Macro, IEEE,
  2008, pp.~552--555.

\bibitem{haupt2009compressive}
{\sc J.~Haupt, R.~Baraniuk, R.~Castro, and R.~Nowak}, {\em {Compressive
  distilled sensing: Sparse recovery using adaptivity in compressive
  measurements}}, in Forty-Third Asilomar Conference on Signals, Systems and
  Computers, IEEE, 2009, pp.~1551--1555.

\bibitem{haupt2009distilled}
{\sc J.~Haupt, R.~Castro, and R.~Nowak}, {\em {Distilled sensing: Selective
  sampling for sparse signal recovery}}, in Proc. 12th International Conference
  on Artificial Intelligence and Statistics (AISTATS), Citeseer, 2009,
  pp.~216--223.

\bibitem{tomographyBook}
{\sc G.~Herman}, {\em {Image Reconstruction from Projections, The Fundamentals
  of Computerized Tomography}}, {New York Academic Press}, 1980.

\bibitem{hoeffding}
{\sc W.~Hoeffding}, {\em Probability inequalities for sums of bounded random
  variables}, J. Amer. Statist. Assoc., 58 (1963), pp.~713--721.

\bibitem{huang1991optical}
{\sc D.~Huang, E.~A. Swanson, C.~P. Lin, J.~S. Schuman, W.~G. Stinson,
  W.~Chang, M.~R. Hee, T.~Flotte, K.~Gregory, C.~A. Puliafito, et~al.}, {\em
  Optical coherence tomography}, Science, 254 (1991), pp.~1178--1181.

\bibitem{clusterAnalysisAstronomy}
{\sc W.~Jang and M.~Hendry}, {\em Cluster analysis of massive datasets in
  astronomy}, Statistics and Computing, 17 (2007), pp.~253--262.

\bibitem{Kaipio.Somersalo.Book2005}
{\sc J.~Kaipio and E.~Somersalo}, {\em Statistical and Computational Inverse
  Problems}, Springer, New York, NY, 2005.

\bibitem{krishnamurthy2011fast}
{\sc K.~Krishnamurthy, W.~Bajwa, R.~Willett, and R.~Calderbank}, {\em Fast
  level set estimation from projection measurements}, in Statistical Signal
  Processing Workshop (SSP), 2011 IEEE, IEEE, 2011, pp.~585--588.

\bibitem{kyrieleisregion}
{\sc A.~Kyrieleis, V.~Titarenko, M.~Ibison, T.~Connolley, and P.~Withers}, {\em
  {Region-of-interest tomography using filtered backprojection: {A}ssessing the
  practical limits}}, Journal of Microscopy,  (2010).

\bibitem{laurent2000adaptive}
{\sc B.~Laurent and P.~Massart}, {\em Adaptive estimation of a quadratic
  functional by model selection}, Annals of Statistics,  (2000),
  pp.~1302--1338.

\bibitem{lewitt1979processing}
{\sc R.~M. Lewitt}, {\em Processing of incomplete measurement data in computed
  tomography}, Medical physics, 6 (1979), p.~412.

\bibitem{li2006fast}
{\sc C.~Li, C.~Xu, K.~Konwar, and M.~Fox}, {\em Fast distance preserving level
  set evolution for medical image segmentation}, in Proc. 9th Intl. Conf.
  Control, Automation, Robotics and Vision, IEEE, 2006, pp.~1--7.

\bibitem{Lu.Do.ITSP2008}
{\sc Y.~Lu and M.~Do}, {\em A theory for sampling signals from a union of
  subspaces}, {IEEE} Trans. Signal Processing, 56 (2008), pp.~2334--2345.

\bibitem{JunMa2010}
{\sc J.~Ma}, {\em Iterative region of interest reconstruction in emission
  tomography}, in IEEE International Symposium on Biomedical Imaging: From Nano
  to Macro,, April 2010, pp.~604 --607.

\bibitem{maa§2011new}
{\sc C.~Maa{\ss}, M.~Knaup, and M.~Kachelrie{\ss}}, {\em New approaches to
  region of interest computed tomography}, Medical Physics, 38 (2011), p.~2868.

\bibitem{marques2009target}
{\sc R.~Marques, F.~De~Medeiros, and D.~Ushizima}, {\em {Target detection in
  SAR images based on a level set approach}}, Systems, Man, and Cybernetics,
  Part C: Applications and Reviews, IEEE Transactions on, 39 (2009),
  pp.~214--222.

\bibitem{mason2009asymptotic}
{\sc D.~Mason and W.~Polonik}, {\em Asymptotic normality of plug-in level set
  estimates}, The Annals of Applied Probability, 19 (2009), pp.~1108--1142.

\bibitem{moreno2003total}
{\sc A.~Moreno}, {\em Total error in a plug-in estimator of level sets},
  Statistics and Econometrics Working Papers,  (2003).

\bibitem{puetter2005digital}
{\sc R.~Puetter, T.~Gosnell, and A.~Yahil}, {\em Digital image reconstruction:
  Deblurring and denoising}, Annual Reviews on Astronomy and Astrophysics, 43
  (2005), pp.~139--194.

\bibitem{rigollet2006fast}
{\sc P.~Rigollet and R.~Vert}, {\em Fast rates for plug-in estimators of
  density level sets}, Bernoulli, 15 (2009), pp.~1154--1178.

\bibitem{rosen2000synthetic}
{\sc P.~Rosen, S.~Hensley, I.~Joughin, F.~Li, S.~Madsen, E.~Rodriguez, and
  R.~Goldstein}, {\em Synthetic aperture radar interferometry}, Proceedings of
  the IEEE, 88 (2000), pp.~333--382.

\bibitem{rudelson2013lecture}
{\sc M.~Rudelson}, {\em Lecture notes on non-asymptotic theory of random
  matrices}, arXiv preprint arXiv:1301.2382,  (2013).

\bibitem{rudin1992nonlinear}
{\sc L.~I. Rudin, S.~Osher, and E.~Fatemi}, {\em Nonlinear total variation
  based noise removal algorithms}, Physica D: Nonlinear Phenomena, 60 (1992),
  pp.~259--268.

\bibitem{scott2007regression}
{\sc C.~Scott and M.~Davenport}, {\em Regression level set estimation via
  cost-sensitive classification}, IEEE Transactions on Signal Processing, 55
  (2007), pp.~2752--2757.

\bibitem{clayAnomaly}
{\sc C.~Scott and R.~Nowak}, {\em Learning minimum volume sets}, Journal of
  Machine Learning Research, 7 (2006), pp.~665--704.

\bibitem{scott2006minimax}
{\sc C.~Scott and R.~Nowak}, {\em Minimax-optimal classification with dyadic
  decision trees}, IEEE Transactions on Information Theory, 52 (2006),
  pp.~1335--1353.

\bibitem{singh2009adaptive}
{\sc A.~Singh, C.~Scott, and R.~Nowak}, {\em {Adaptive Hausdorff estimation of
  density level sets}}, The Annals of Statistics, 37 (2009), pp.~2760--2782.

\bibitem{riceCamera}
{\sc D.~Takhar, J.~Laska, M.~Wakin, M.~Duarte, D.~Baron, S.~Sarvotham,
  K.~Kelly, and R.~Baraniuk}, {\em A new compressive imaging camera
  architecture using optical-domain compression}, in SPIE, vol.~6065, 2006.

\bibitem{tibshirani1996regression}
{\sc R.~Tibshirani}, {\em Regression shrinkage and selection via the lasso},
  Journal of the Royal Statistical Society. Series B (Methodological),  (1996),
  pp.~267--288.

\bibitem{matchingPursuit}
{\sc J.~Tropp and A.~Gilbert}, {\em Signal recovery from random measurements
  via orthogonal matching pursuit}, Information Theory, IEEE Transactions on,
  53 (2007), pp.~4655 --4666.

\bibitem{tsybakovDensity}
{\sc A.~Tsybakov}, {\em On nonparametric estimation of density level sets},
  Annals of Statistics, 25 (1997), pp.~948--969.

\bibitem{vaiter2011robust}
{\sc S.~Vaiter, G.~Peyr{\'e}, C.~Dossal, and J.~Fadili}, {\em Robust sparse
  analysis regularization}, Arxiv preprint arXiv:1109.6222,  (2011).

\bibitem{vapnik2000nature}
{\sc V.~Vapnik}, {\em The nature of statistical learning theory}, Springer
  Verlag, 2000.

\bibitem{Vershynin_notes}
{\sc R.~Vershynin}, {\em Norm of a random matrix}.
\newblock Lecture notes on \emph{Non-asymptotic theory of random matrices},
  2006-2007.

\bibitem{wang2008new}
{\sc Y.~Wang, J.~Yang, W.~Yin, and Y.~Zhang}, {\em A new alternating
  minimization algorithm for total variation image reconstruction}, SIAM
  Journal on Imaging Sciences, 1 (2008), pp.~248--272.

\bibitem{willett:levelset}
{\sc R.~Willett and R.~Nowak}, {\em Minimax optimal level set estimation}, IEEE
  Transactions on Image Processing,  (2007), pp.~2965--2979.

\bibitem{yang}
{\sc Y.~Yang}, {\em {Minimax nonparametric classification-Part I: Rates of
  convergence}}, IEEE. Transactions on Information Theory, 45 (1979),
  pp.~2271--2284.

\end{thebibliography}
\end{document}